\documentclass[journal]{IEEEtran}

\usepackage{cite}
\newcommand{\etal}{\textit{~et~al.~}}
\usepackage[pdftex]{graphicx}
\usepackage{subfig}
\graphicspath{{../pdf/}{../jpeg/}}
\DeclareGraphicsExtensions{.pdf,.jpeg,.png}
\usepackage{amsmath}
\usepackage{amssymb}
\newcommand{\norm}[1]{\left\lVert#1\right\rVert}
\DeclareMathOperator{\sgn}{sgn}
\DeclareMathOperator{\range}{Range}
\DeclareMathOperator{\tr}{Tr}
\usepackage{array}
\usepackage{color}

\DeclareMathOperator{\diag}{diag}
\usepackage{makecell}
\usepackage{amsthm}
\usepackage{eqnarray}
\newtheorem{theorem}{Theorem}

\newtheorem{lemma}[theorem]{Lemma}
\newtheorem{proposition}[theorem]{Proposition}
\newtheorem{definition}[theorem]{Definition}
\begin{document}

\title{Separable Joint Blind Deconvolution and Demixing}
\author{Dana~Weitzner
        and~Raja~Giryes% <-this % stops a space
% \thanks{D.Weitzner and R.Giryes are with the School of Electrical Engineering, Faculty of Engineering, Tel-Aviv University, Ramat Aviv 69978, Israel.  e-mail:~\{danaweitzner@mail, raja@tauex\}.tau.ac.il.}% <-this % stops a space
% \thanks{Manuscript received April 19, 2005; revised August 26, 2015.}
\thanks{\copyright 2021 IEEE.  Personal use of this material is permitted.  Permission from IEEE must be obtained for all other uses, in any current or future media, including reprinting/republishing this material for advertising or promotional purposes, creating new collective works, for resale or redistribution to servers or lists, or reuse of any copyrighted component of this work in other works.}
}

% The paper headers
% \markboth{IEEE JOURNAL OF SELECTED TOPICS IN SIGNAL PROCESSING}%
% {Weitzner \MakeLowercase{\textit{et al.}}: Separable Joint Blind Deconvolution and Demixing}
% make the title area
\maketitle

\begin{abstract}
Blind deconvolution and demixing is the problem of reconstructing convolved signals and kernels from the sum of their convolutions. This problem arises in many applications, such as blind MIMO. 
This work presents a separable approach to blind deconvolution and demixing via convex optimization. Unlike previous works, our formulation allows separation into smaller optimization problems, which significantly improves complexity. We develop recovery guarantees, which comply with those of the original non-separable problem, and demonstrate the method performance under several normalization constraints. 
\end{abstract}

% Note that keywords are not normally used for peer-review papers.

\section{Introduction}\label{sec:introduction}
\IEEEPARstart{C}{onsider} the task of restoring signals from a mixture of their bilinear measurements, involving unknown environment parameters. This problem is referred to as blind deconvolution and demixing, where signals are convolved with unknown kernels. It appears in various domains, e.g., audio and image processing \cite{Audio_processing, image_processing_1, image_processing_2} and wireless communications \cite{wireless_communication}, in which it is expected to play a central role in IoT \cite{IoT}.

In the problem of joint deconvolution and demixing \cite{blind_deconv_demix}, the goal is to  reconstruct the signals $x_s$ and kernels $w_s$ from
\begin{equation}\label{eq:convolutions_sum}
y =\sum_{s \in [S]} x_s \circledast w_s,
\end{equation}
extending blind deconvolution to a sum of convolutions. 
Like the classic blind deconvolution problem \cite{blind_deconvolution}, this problem is ambiguous without further constraints on the signals and kernels (more on the ambiguities of one dimensional blind deconvolution can be found in \cite{blind_deconv_ambiguities, blind_deconv_ambiguities_2}). Common assumptions include peakiness (e.g., \cite{peaky_signals_sparse_power_factorization_method, peaky_signals_sparse_power_factorization_theory}),  sparsity (e.g., \cite{nonconvex_sparse_and_low_rank, sparsity_prior}),  and subspace priors (e.g., \cite{subspace_prior_blind_deconv}).

Our problem \eqref{eq:convolutions_sum} was solved using convex Nuclear norm minimization, exploiting the rank-1 structure of the lifted problem \cite{PhaseLift, Exact_Matrix_Completion, blind_deconv_demix, linear_blind_deconv_demix_one_measurement}, assuming the subspace prior suggested by \cite{subspace_prior_blind_deconv}. Probabilistic linear guarantees for the relationship between the amount of measurements in $y$, and the amount of signal-kernel pairs, $S$, were derived for this method \cite{linear_blind_deconv_demix_one_measurement}.  

The convex Nuclear minimization approach allows the derivation of theoretical guarantees with minimal assumptions, though combined with the common lifting procedure it might result in high computational complexity. Non-convex approaches were also explored in the context of blind deconvolution \cite{Nonconvex_deconv} and demixing \cite{Nonconvex_deconv_demix, Nonconvex_deconv_demix_2, nonconvex_hard_thresholding_quadratic, nonconvex_Riemannian_quadratic, nonconvex_quadratic_WF_randm_init, nonconvex_Wirtinger_Blair_quadratic}, with significant lower computation time. Though the theoretical result in \cite{Nonconvex_deconv} is in line with those achieved in the convex approach, the expansions to demixing via the non-convex methods are still with quadratic guarantees.
A thorough review of nonconvex algorithms in a broader context of general matrix completion problems can be found in \cite{nonconvex_matrix_factorization_review}.

A related variant of \eqref{eq:convolutions_sum} considers a scenario where $S$ sources transmit signals to $R$ receivers. Each path is modeled by a different, unknown convolution kernel, yielding
\begin{equation} 
y_r =\sum_{s \in [S]} x_s \circledast w_{rs},
 \qquad   r \in [R],
\end{equation}
where $y_r$ is the measurement of the $r$th out of $R$ receivers. This is equivalent to the blind MIMO model presented in \cite{blind_mimo_AA}. Their method also uses Nuclear norm minimization employing the rank-1 structure of the lifted problem, assuming that the signals reside in a known subspace. This leads to an optimization problem on a matrix consisting of rank-1 blocks, which do not share variables. However, the solution cannot be obtained separately for each block and requires solving the full optimization problem of dimension $RK\times SN$, where $K, N$ are the dimensions of the signals and kernels subspaces.  

{\bf Contribution.} This work assumes that all signals lay in the same subspace (as do all kernels). This allows solving separate rank-$S$ problems instead of a single, large rank-$1$ problem, which reduces the computational complexity significantly. We show linear performance in the reconstruction of each rank-$S$ matrix and develop theoretical guarantees that match those of the rank-1 case \cite{linear_blind_deconv_demix_one_measurement}.
The advantage of our solution is its improved computational complexity; Instead of a single large problem, we solve a few small optimization problems with better variables to degrees of freedom ratio.  See Section \ref{sec:method}. % elaborates on the complexity analysis. 

Given the recovered rank-$S$ matrix, the standard form to retrieve the originating vectors from it is by SVD~\cite{subspace_prior_blind_deconv}. However, in the rank-$S$ case, there is an ambiguity of the spanning base, i.e., the singular vectors are generally not the originating vectors. To overcome this, we suggest an algorithm that uses the fact that the signals are shared between all receivers, and can reconstruct all signals and kernels under some normalization assumptions, to be described in Section \ref{sec:basis_trans}. 

% Uniqueness is assured trivially in the exponential case, i.e., when $R=2^{S-1}$. However, numerical simulations show that under our assumptions, the unique solution is obtained by our method when $R=S$ and both signals and kernels are normalized, or $R=S+1$ when only the kernels are normalized. This additional step to resolve the spanning base ambiguity preserves the linear performance of the original rank-1 formulation, as we show hereafter. 

{\bf Notations.}
Unless stated otherwise, we use the following norm notations throughout our derivations. $\norm{\cdot}$ without subscript is the operator norm of the appropriate subject: $ \norm{\mathcal{A}} = \norm{\mathcal{A}}_{F\to2}$ for sampling operators; $ \norm{A} = \norm{A}_{2\to2}$ for matrices. We use $[S]$ to describe the set of integers $1,\dots, S$. 
$x_s$/$y_{rs}$ denotes the $s$ column of a matrix $X$/$Y_r$. To denote an inequality up to a constant depending on $\omega$ we use $\lesssim_\omega$.

% \section{Demixing convolved signals and kernels}\label{sec:general_model}

\section{The problem setup}\label{sec:model}

Consider that $R$ convolutions-sums are measured, sharing the signals while the kernels are different per source-receiver path. Let the convolution kernel matrix for each sensor be $W_r \in \mathbb{C}^{L \times S}$, and let $X \in \mathbb{C}^{L \times S}$ be $S$ signals of length $L$. Assume that the signals and kernels reside in a low $(N, K)$ dimensional subspaces spanned by the columns of the known matrices $B$, $C$, such that 
\begin{eqnarray}
& X=CM, &  M \in \mathbb{C}^{N \times S},\\
& W_r=BH_r, &  H_r \in \mathbb{C}^{K \times S},
\end{eqnarray}
where $B \in \mathbb{C}^{L \times K}$ is assumed to have orthogonal columns, while the entries of $C \in \mathbb{C}^{L \times N}$ are independent and follow a standard circular-symmetric normal distribution. The standard subspace prior \cite{subspace_prior_blind_deconv} can be achieved in our framework with an appropriate choice of coding matrices. Therefore, our assumption is reasonable and holds no limitation to the common approach. Denote the "column-wise" cyclic convolution of $X$ and $W_r$ as $X \circledast W_r = \tilde{Y}_r \in \mathbb{C}^{L \times S}$. Then
the measurement of each sensor reads as
\begin{equation}\label{eq:measurement_with_noise}
y_r =\sum_s x_s \circledast w_{rs} + e = \sum_s \tilde{y}_{rs} + e,
 \qquad  s\in [S],
\end{equation}
where $e$ is additive noise. Let $F$ be the $L$ dimensional DFT matrix, $\diag(f_l)$ the diagonal matrix consisting of the $l$th column of $F$, $f_l$, and $A_l = \sqrt{L} B^* F^* \diag(f_l) \bar{F} \bar{C}$. Then for a given signal-kernel pair, the $l$th measurement component in Fourier domain is given by \cite{Bresler}
\begin{equation} 
\hat{\tilde{y}}_{rsl} \triangleq (F(x_s \circledast w_{rs}))_l = \langle A_l , m_s h^*_{rs} \rangle, 
\end{equation}
Thus, the $l$th Fourier entry of the signal-kernel pairs sum is
\begin{eqnarray} 
\hat{y}_{rl}  &\triangleq & (F(y_r))_l  = \sum_s \hat{\tilde{y}}_{rsl} + \hat{e}_l
 = \sum_s \langle A_l , m_s h_{rs}^* \rangle + \hat{e}_l \nonumber\\
&=& \langle A_l , \sum_s m_s h_{rs}^* \rangle + \hat{e}_l
 = \langle A_l , MH^*_r \rangle + \hat{e}_l.
\end{eqnarray}
The complete linear measurement operator $\mathcal{A}:\mathbb{C}^{N\times K} \rightarrow \mathbb{C}^L$ in Fourier domain is therefore  defined by
\begin{equation}\label{A}
\mathcal{A}(\cdot)=[\langle A_1 , \cdot \rangle ,\ldots, \langle A_L , \cdot \rangle]^T,
\end{equation}
which leads to writing the measured vector at receiver $r$ as
\begin{equation} 
\hat{y}_r = \mathcal{A}(MH^*_r) + \hat{e} = \mathcal{A}(Z_r) + \hat{e}, 
\label{eq:linear_measurement}
\end{equation}
where $Z_r \triangleq  MH^*_r$. The problem of demixing convolved signals and kernels is hence the reconstruction of the signals subspace coefficients vectors $M$ and the convolution kernels subspace coefficients vectors $H_r$ from the Fourier transform of the measurement vector $y_r$. 

\section{Separable optimization for SVD based joint deconvolution and demixing}\label{sec:method}
Casting the problem as a matrix recovery problem, as in \eqref{eq:linear_measurement}, allows the use of rank minimization algorithms since $Z_r$ is known to be a rank-$S$ matrix. Yet, unlike the rank-$1$ case, the recovery of the matrix is insufficient for the reconstruction of the actual signals and kernels. As we further explain hereafter, this is due to a wider ambiguity in the factorization of the matrix, that is not resolved in the SVD process, which is the standard tool for vector recovery in the rank-$1$ case \cite{subspace_prior_blind_deconv}. 

Our framework considers a model in which $S$ sources transmit signals to $R$ receivers, while each channel is represented by a different convolution kernel. Thus, each receiver measures the mixture of $S$ convolutions, $y_r$. Our method has two stages: 

\noindent 1. \emph{Matrix recovery}: reconstruct $Z_r$ from $y_r$ via Nuclear norm minimization at each receiver, separately.

\noindent 2. \emph{Vector recovery}: estimate $M, H_r$ from $Z_r = MH^*_r$. This step uses the estimated $Z_r$ from all receivers and requires solving a quadratic equation system of $S^2$ variables, regardless of $L, K$, and $N$, which may be generally much larger.

\subsection{Matrix recovery}\label{sec:matrix_rec}
Assume that all signals and kernels have the same coding matrix. Thus, for each receiver, we may recover $Z_r \in  \mathbb{C}^{K \times N}$, which has \textbf{rank S}, by  solving 
\begin{equation}\label{optimization_probelm}
 \min_Z    \norm{Z}_* \qquad  \text{s.t.} \qquad \norm{\mathcal{A}(Z) - \hat{y}_r}_2 \leq \tau.
\end{equation}
Computationally, this is equivalent to the standard convex approach to blind deconvolution, with no demixing \cite{subspace_prior_blind_deconv}.

For comparison, \cite{linear_blind_deconv_demix_one_measurement} deal only with the case of one receiver ($R=1$) so the total degrees of freedom (DoF) are $S (K + N)$. In the noiseless case, they solve an optimization problem for $S$ rank-1 matrices, i.e. with $SKN$ variables:
\begin{equation}\label{eq:big_problem}
         \min_{Z_s} \sum_s \norm{Z_s}_*  \qquad \text{s.t.}\qquad \norm{\sum_s \mathcal{A}_s(Z_s) - \hat{y}}_2 \leq \tau.
\end{equation}
Compared to this approach, we have more measurements per signal ($R\geq S$); This trades off the number of measurements and the ability to solve smaller problems.

Ahmed \cite{blind_mimo_AA} presented a convex approach to blind MIMO, which is more similar to our case in the sense that he too has multiple receivers. Yet, he solves the problem directly (i.e. with rank one blocks) so he may have $R < S$. When $R=1$, this is exactly \eqref{eq:big_problem}. When $R \geq S$, this is equivalent to our problem. It has $SN + RSK$ degrees of freedom ($S$ signals of length $N$ and $S$ kernels of length $K$ in each of the $R$ receivers, without the normalization implications). He solves a similar problem to the former case, only for a sum of $S$ rank-1 matrices of size $RK \times N$, meaning with $SRKN$ variables. As they also note, although the different matrices in the sum share no common variables, the problem cannot be separated. 

In our setup we use the shared information to be able to separate the problems, but then the information is no longer shared and there is no sample complexity gain. The "full" approach of the same scheme \cite{blind_mimo_AA}, is on the other side of the trade-off - they show better sample complexity for more receivers but solve a bigger problem. Note though that they show it only empirically, where the dependency decreases quite quickly with the increase in $R$. They also conjecture a linear sample complexity bound (which is what we formally prove). It would be interesting to think of an approach in the middle of the trade-off, where we use our shared subspace assumption in a way that has provably better sample complexity. We leave this to future work.

Table~\ref{tab_matrix_rec_comp} summarizes the complexity of each method. For each of the methods mentioned above, we detail the desired DoFs, the actual number of variables in the problem, the number of sub-problems ("\#Prob"), extra optimization variables ("ext. op."), the total number of variables, and finally the ratio between the optimization variables and the desired DoF. The bottom line includes the range of receivers regime handled by each method. Notice that we have fewer optimization variables, as we keep our matrices small. 
Moreover, we also have a better (lower) ratio between the optimization variables and the degrees of freedom. 

\begin{table}
\caption{Complexity comparison of the different methods. \#Prob. is the number of separate problems in each work.}
\label{tab_matrix_rec_comp}
\centering
\begin{tabular}{|c|c|c|c|}
\hline
\bfseries Method & \cite{linear_blind_deconv_demix_one_measurement} & 
\cite{blind_mimo_AA} & Ours \\ \hline \hline
\bfseries DoF & $S(K+N)$ & $S(RK+N)$ & $S(RK+N)$ \\ \hline
\bfseries \makecell{Opt. \\vars} & $SKN$ & $SRKN$ & $KN$ \\ \hline
\bfseries \makecell{\#Prob.} & 1 & 1 & $R$ \\ \hline
\bfseries \makecell{ext. op.} & - & - & $S^2$ \\ \hline
\bfseries Total & $SKN$ & $SRKN$ & $R\times KN+S^2$ \\ \hline
\bfseries Ratio & $\frac{KN}{K+N}$ & $\frac{RKN}{RK+N}$ & $\frac{RKN+S^2}{S(RK+N)}$ \\ \hline
\bfseries R & $R=1$ & $R\geq1$ & $R \geq S$ \\ \hline
\end{tabular}
\end{table}

\subsection{Vector recovery: find basis transformation}\label{sec:basis_trans}

Once $Z_r$ is restored, we use its SVD decomposition to reconstruct the originating vectors (the signals and kernels coefficients). Thus $Z_r = U \Lambda V^* \triangleq \tilde{M}_r \tilde{H}_r^*$, where $\tilde{m}_{ri} = \sqrt{\Lambda_{ii}}u_i$ (similarly for $\tilde{h}_{ri}$). 
In the rank-1 case, this is trivial: 
$Z_r = m h_r^* = \tilde{m}_{r1} \sigma \sigma^{-1} \tilde{h}_{r1}^*$ for any $\sigma \neq 0$.
Thus, the estimated vectors are $m=\tilde{m}_{r1}, h_r=\tilde{h}_{r1}$ up to scale and sign. 
These ambiguities of the rank-1 case become an ambiguity of spanning base in higher ranks, which can be expressed by 
\begin{equation}\label{eq:rank-R_ambiguity}
   MH_r^* = \tilde{M}_r T^r (T^r)^{-1} \tilde{H}_r^*, 
\end{equation}
where $T^r \in \mathbb{C}^{S \times S}$ is the basis transformation from the columns of $\tilde{M}_r$ to the columns of the original $M$. 

To restore the original vector pairs, we measure $R$ convolutions-sums, assuming that $M$ is constant in all of them while only $H_r$ is changing. This corresponds to the blind MIMO scheme in \cite{blind_mimo_AA} and allows us to pose enough constraints to determine $T^r, \forall r\in [R]$, as follows:
To exploit the fact that $M$ is shared, we want to express the relation in \eqref{eq:rank-R_ambiguity} with the same transformation matrix $T$, for all $r\in[R]$. Thus, we choose an arbitrary $r_0\in [R]$. The relation between the transformation matrices at different receivers is given by
\begin{equation}
     T^r = (T^{rr_0})^{-1}T^{r_0},
\end{equation}
%where
%\begin{equation}
%\end{equation}
where $T^{rr_0} \triangleq \tilde{M}_{r_0}^+ \tilde{M}_r$ and $ ^+$ is the pseudo-inverse. Note that we can always invert $\tilde{M}_{r_0}^* \tilde{M}_{r_0}$ as $\tilde{M}_{r_0}$ has orthogonal columns (due to SVD).  
To recover the original coefficients, it is sufficient to find the $S^2$ entries of $T^{r_0}$. Note that it satisfies
\begin{align} 
M &= \overbrace{\tilde{M}_r (T^{rr_0})^{-1}}^{\tilde{M}_{r_0}} T^{r_0},\label{eq:Qr} \\
H_r &= \underbrace{\tilde{H}_r (T^{rr_0})^{*}}_{\triangleq G_r} (T^{r_0})^{-*}.\label{eq:Gr}
\end{align}
This equations system allows adding constraints on the same variable basis transformation matrix, $T^{r_0}$, by increasing $R$. Note that we do not know $M$ and $H_r$. Thus, we need to add some constraints on them to be able to recover $T^{r_0}$.
%These constraints resolve the spanning base ambiguity and preserve the linear performance of the original rank-1 formulation, as we show hereafter. 
%We next examine two options. 

{\bf One-sided constraints.}
Assuming normalized kernel coefficients (a standard assumption due to the scale ambiguity of rank-1 case) and considering \eqref{eq:Qr} leads to the equations
\begin{equation}
\label{eq:one_sided_eq}
% \norm{m_s}^2 = 1 = (T^{r_0})_s^* Q_r^* Q_r T^{r_0}_s, \qquad 
\norm{h_{rs}}^2 = 1 = ((T^{r_0})^{-*}_s)^* G_r^* G_r  (T^{r_0})^{-*}_s, r\in [R],
\end{equation}
where $T^{r_0}_s$ is the $s$th column of $T^{r_0}$. Having the original coefficients vectors equally normalized means that the equation system is column separable w.r.t $T^{r_0}$. Thus, we get $R$ equations for the $S$ variables of $T^{r_0}_s$ for any $s\in [S]$. 

Posing $l_2$ norm constraints on the original signals or kernels coefficients leads to a system of \textbf{quadratic} equations, which has an exponential number of solutions even in the fully determined case; A fully determined system with $S$ equations and $S$ variables lead to $2^{S}$ solutions. When $R=S$, the system has at least $2^{S-1}$ solutions (we are agnostic to global phase ambiguity). While this might suffice when $S=2$, it is insufficient for more signal-kernel pairs. Empirically we observed that solving this equation system by optimization for the entire matrix $T^{r_0}$, i.e., simultaneously solving for all columns of $T^{r_0}$,  produces only the $S$ correct results even for $R = S +1$, despite the exponential amount of valid solutions. This was resolved using the Matlab non-linear equation solver fsolve with the trust region Dogleg algorithm.

{\bf Two-sided constraints.}
Posing the normalization constraints on both $M$ and $H_r$, and using it with \eqref{eq:Qr} and \eqref{eq:Gr}, leads to the equation system  
\begin{eqnarray}
           \label{eq:first_subproblem}
            \norm{m_s}^2 = 1 =& (T^{r_0}_s)^* \tilde{M}_{r_0}^* \tilde{M}_{r_0} T^{r_0}_s,\\
            \norm{h_{rs}}^2 = 1 =& ((T^{r_0})^{-*}_s)^* G_r^* G_r (T^{r_0})^{-*}_s. \label{eq:second_subproblem}
\end{eqnarray}
This adds an additional constraint. However, although each subproblem (\eqref{eq:first_subproblem}, \eqref{eq:second_subproblem}) is column separable, the total system is not, as the variables in each of the separate sets are the adjugate matrices of one another (up to the determinant factor). In fact, we are looking for the correct set of $S$ solutions, out of $\binom{2^{S-1}}{S}$ possible column sets that can solve \eqref{eq:first_subproblem}, which can be jointly inverted to the correct set of solutions of \eqref{eq:second_subproblem}. This setup appears to impose "hidden" constraints. We conjecture that they resolve the ambiguity of the quadratic equations for $R=S$, as we empirically show in Section \ref{sec:experiments}.

% \subsubsection{$l_1$ norm constraints?}
% It can be algebraically shown that equations are the same for the linear setup. 

\subsection{Matrix recovery guarantees} 
We turn now to provide theoretical guarantees for the recovery of $Z_r$.
Jung\etal \cite{linear_blind_deconv_demix_one_measurement} were the first to present a linear guarantee for the uniqueness of recovery when solving \eqref{eq:big_problem}. Ahmed \cite{blind_mimo_AA}, which presented a convex approach to blind MIMO, conjectured that also in his case $L$ is linear w.r.t. $\max(R,S)$, without proof. Such guarantees \cite{subspace_prior_blind_deconv, blind_deconv_demix, linear_blind_deconv_demix_one_measurement} rely on the fact that the retrieved matrix is \textbf{rank-1}, and, thus, do not apply in our case. The following provides a guarantee for the reconstruction of higher rank matrices (by solving \eqref{optimization_probelm}), with the same assumptions on $B, C$ and a similar linear result. 
% B - incoherent in the Fourier domain, orthogonal columns (fits our example). define all types of mu - raise the point that all proofs for the uniqueness of matrix retrieval assume our constraints for the decomposition process.
% C - generic, independent rows. explain concept of proof. 
\begin{theorem}\label{main_theorem}
Let $\omega \geq 1$ and let $y \in \mathbb{C}^L$ be given by \eqref{eq:measurement_with_noise}, with $\norm{e}_2 \leq \tau$. Assume that 
\begin{equation}
     L \gtrsim_\omega  S (K\mu^2 \log(K\mu^2) + N\mu^2_H) \log^3L,  
\end{equation}
then with probability of at least $1-\mathcal{O}(L^{-\omega})$ the minimizer $\hat{X}$ of \eqref{optimization_probelm} satisfies
\begin{equation}
    \norm{\hat{X} - X_0}_F \lesssim_\omega \tau \sqrt{S \max \bigg{\{}1, \frac{SK\mu^2N}{L}\bigg{\}}\log^3 (L)}. 
\end{equation}
\end{theorem}

Compared to \cite{linear_blind_deconv_demix_one_measurement}, we have the same lower bound on $L$ but effectively more measurements per signal (all $y$-vectors). This is part of the tradeoff that enables us to have smaller computational problems, as discussed in Section~\ref{sec:matrix_rec}.
\section{Proof of the main theorem}

The structure of our proof is similar to the ones in \cite{subspace_prior_blind_deconv, blind_deconv_demix, linear_blind_deconv_demix_one_measurement}. We prove sufficient conditions for recovery, assuming the existence of an inexact dual certificate: We show that our measurement operator fulfills a Local Isometry Property (see \eqref{LIP_T}) on the relevant spaces (those defined in Def. \ref{solution_space} and \ref{extended_solution_space}), and then construct the dual certificate using the Golfing Scheme.

\subsection{Preliminary Definitions}

We start with preliminary definitions. The $\sgn$ function is defined in the functional sense, i.e.
\begin{equation}\label{def:sgn_func}
    \sgn(A) = U \text{diag}(\sgn(\sigma_1), \dots , \sgn(\sigma_r)) V^*,
\end{equation}
where $A=U\Sigma V^*$ is the SVD of $A$ and $\{\sigma_i\}$ are the singular values of $A$. This differs from the definition in \cite{linear_blind_deconv_demix_one_measurement} and is important for proving the results for the reconstruction of matrices with rank exceeding one.
We denote by  $\hat{H}\Lambda\hat{M}^*$ the SVD decomposition of $X_0=HM^*$, and define 
\begin{equation}\label{M}
    \mathcal{M} = \big{\{} Z \ |\ Z \in \mathbb{C}^{K\times N} \big{\}},
\end{equation}
to be the space of matrices of the appropriate size.

\textbf{Solution space.} We now turn to define the solution space. (Note that some other works refer to it as the tangent space.)
\begin{definition}[Solution space]\label{solution_space}
Let $\hat{H}\Lambda\hat{M}^*$ be the SVD decomposition of $X_0=HM^*$. Given
\begin{equation*}%\label{T_M}
    \mathcal{T}_M = \big{\{} V\hat{M}^* \ |\ V \in \mathbb{C}^{K\times S} \big{\}},
    \mathcal{T}_H = \big{\{} \hat{H}U^* \ |\ U \in \mathbb{C}^{N\times S} \big{\}},
\end{equation*}
% \begin{equation}\label{T_H}
%     \mathcal{T}_H = \big{\{} \hat{H}U^* \ |\ U \in \mathbb{C}^{N\times S} \big{\}}
% \end{equation}
the solution space is defined as $\mathcal{T} = \mathcal{T}_M +\mathcal{T}_H$.
\end{definition}

\textbf{Local Isometry Property (LIP).} An operator $\mathcal{A}$ satisfies the LIP with a constant $\delta$ if $\forall X\in \mathcal{T}$
\begin{equation}\label{LIP_T}
    (1-\delta)\norm{X}_F^2 \leq \norm{\mathcal{A}(X)}_2^2 \leq (1+\delta)\norm{X}_F^2.
\end{equation}

\textbf{Partition of measurements and incoherence.} In our proofs, we also use extended spaces that are slightly larger than the solution space. These spaces are induced by the measurements partitioning, required for the Golfing scheme \cite{Golfing_Scheme}.
Define the kernels subspace matrix coherence parameter
\begin{equation}\label{B_incoherence}
    \mu^2 = \frac{L}{K} \underset{l\in[L]}{\max} \norm{b_l}_2^2,
\end{equation}
where $b_l$ is the $l$th column of $B^T$. Notice that $1 \leq \mu^2 \leq \frac{L}{K}$.

Using the Golfing Scheme~\cite{Golfing_Scheme} requires a division of the $L$ measurements into $P$ non-overlapping sets. We denote the indexing of each set by $\Gamma_p$ where $p\in[P]$. Thus, $\cup_p \Gamma_p = [L]$. Each set is associated with its linear measurement operator
\begin{equation}\label{A_p}
    \mathcal{A}^p(Z) = \{ \langle A_l, Z \rangle \}_{l \in \Gamma_p}.
\end{equation}
For convenience of writing, we define
\begin{eqnarray}\label{T_p}
    & T_p = \frac{L}{Q}\sum_{l\in \Gamma_P} b_l b_l^*, \\
\label{S_p}
    & S_p = T_p^{-1} = \bigg{(} \frac{L}{Q}\sum_{l\in \Gamma_P} b_l b_l^* \bigg{)} ^{-1},
\end{eqnarray}
where $Q \triangleq \frac{L}{P}$. To guarantee the convergence of the Golfing Scheme with high probability, the partition must be chosen such that $T_p\approx I_K$ for all $p\in[P]$. Thus, we require that
\begin{equation}\label{eq:T_p_I}
    \underset{p\in[P]}{\max} \norm{T_p - I} \leq \nu,
\end{equation}
for a small enough $\nu$. 
Moreover, the partition needs to be $\omega$ admissible in the following sense.
\begin{definition}\label{w_admissible}
Let $\omega \geq 1$ and let $\{\Gamma_p\}_{p\in P}$ be a partition of $[L]$. The set $\{\Gamma_p\}_{p\in P}$ is said to be $\omega$-admissible if the following conditions are satisfied:
 \begin{enumerate}
     \item $\frac{1}{2}Q \leq | \Gamma_p | \leq \frac{3}{2}Q$ for all $p \in P$, where $Q=\frac{L}{P}$
     \item  \eqref{eq:T_p_I} is fulfilled with $\nu = \frac{1}{32}$
     \item $\frac{1}{2} \log(8\tilde{\gamma} \sqrt{S}) \leq P \leq \log(8\tilde{\gamma} \sqrt{S})$, where 
     \begin{equation*}
         \tilde{\gamma} = 2\sqrt{\omega \ \max \big{\{} 1, \frac{SK\mu^2 N}{L}\big{\}} \log (L+SKN)}.
     \end{equation*}
 \end{enumerate}
\end{definition}
The existence of such partition is guaranteed by Lemma 3 in \cite{linear_blind_deconv_demix_one_measurement}. 
For a fixed $\omega$-admissible partition we can define 
\begin{equation}\label{H_incoherence}
    \mu_H^2 = L \max \big{\{} \underset{l \in L, s \in S}{\max} |b_l^*h_s|^2, \underset{l \in L, s \in S, p \in P}{\max} |b_l^* S_p h_s|^2 \big{\}},
\end{equation}
where $1\leq \mu_H^2 \leq (\frac{32}{31})^2 K\mu^2 \lesssim L$. Now we can define the extended solution space for each $p$. 
\begin{definition}[Extended solution space]\label{extended_solution_space}
Fix $p\in[P]$. The extended solution space is defined as $\mathcal{T}^p = \mathcal{T} +\mathcal{T}_{S^pH}$, where
\begin{equation}\label{T_S_p}
    \mathcal{T}_{S^pH} = \big{\{} S_p \hat{H}U^* \ |\ U \in \mathbb{C}^{N\times S} \big{\}}.
\end{equation}
\end{definition}

\textbf{Orthogonal projection operators.} We can define the orthogonal projection operator onto the solution space $\mathcal{P}_\mathcal{T}$ by
\begin{equation}\label{P_T}
    \mathcal{P}_\mathcal{T}(Z) = \mathcal{P}_{\hat{H}} Z +Z \mathcal{P}_{\hat{M}} -\mathcal{P}_{\hat{H}} Z \mathcal{P}_{\hat{M}}
\end{equation}
and the orthogonal projection operator onto the complementary space $\mathcal{T}^\perp$ by
\begin{equation}\label{P_T_perp}
    \mathcal{P}_{\mathcal{T}^\perp}(Z) = (I - \mathcal{P}_{\mathcal{T}}) Z = (I_K - \mathcal{P}_{\hat{H}}) Z (I_N - \mathcal{P}_{\hat{M}}),
\end{equation}
where
\begin{eqnarray}\label{vec_proj}
    && \hspace{-0.6in} \mathcal{P}_{\hat{H}} = \hat{H}\hat{H}^*,
    \mathcal{P}_{\hat{M}} = \hat{M}\hat{M}^*, \\
\label{T_perp}
&&   \hspace{-0.6in}   \mathcal{T}^\perp = \text{span}\{ vu^*  \ | \ u\perp\{h_s\}_{s\in[S]}, v\perp\{m_s\}_{s\in[S]} \}.
\end{eqnarray}

\subsection{Sufficient conditions for recovery}

% \begin{lemma}[Uniqueness in the noiseless case]\label{sufficient_condition_for_recovery_noiseless_case}
% Suppose that $\mathcal{A}$ satisfies the $\delta$-local isometry property on $\mathcal{T}$ and set $\gamma = \norm{\mathcal{A}}$, i.e., $\gamma$ is the operator norm of $\mathcal{A}$. Furthermore, suppose that there is $Y \in\range(\mathcal{A^*})$ such that
% \begin{equation}
%     \norm{\mathcal{P}_\mathcal{T} (Y) - \sgn (X_0)}_F \leq \frac{1}{2l^2}
% \end{equation}
% \begin{equation}
%     \norm{\mathcal{P}_{\mathcal{T}^\perp} (Y)} \leq \frac{1}{2}
% \end{equation}
% $\sgn(x) = x/|x|$ in terms of functional analysis (i.e., operating on the eigenvalues). Then $X_0$ is the unique minimizer of Problem \ref{optimization_probelm} with $\tau = 0$.
% \end{lemma}
We first find sufficient conditions for recovery in the presence of noise.
\begin{lemma}\label{sufficient_condition_for_recovery}
Suppose that $\mathcal{A}$ satisfies the $\delta$-local isometry property on $\mathcal{T}$ and set $\gamma = \norm{\mathcal{A}}$. Furthermore, suppose that there is $Y = \mathcal{A}^*z$ for some $z \in \mathbb{C}^L$ such that
\begin{eqnarray}\label{eq:alpha}
    & \norm{\mathcal{P}_\mathcal{T} (Y) - \sgn(X_0)}_F \leq \alpha \\ 
\label{eq:beta}
    & \norm{\mathcal{P}_{\mathcal{T}^\perp} (Y)} \leq \beta
\end{eqnarray}
where $\alpha, \beta \geq 0$ are constants such that $1 -\beta -\frac{\alpha\gamma}{\sqrt{1-\delta}} \geq \frac{1}{2}$, $\alpha \leq 1$ and $\sqrt{1-\delta}\geq\frac{1}{2}$. If $\hat{X}$ is a minimizer of 
% \begin{eqnarray}
%     &\min &\norm{X}_*\\
%     &\text{subject to} &\norm{\mathcal{A}(X) - \hat{y}}_2 \leq \tau.
% \end{eqnarray}
\begin{equation}\label{eq:suf_cond_min_prob}
     \min_X \norm{X}_* \qquad 
     \text{s.t} \qquad \norm{\mathcal{A}(X) - \hat{y}}_2 \leq \tau,
\end{equation}
then 
\begin{equation}\label{X_err_bound}
    \norm{\hat{X} - X_0}_F \lesssim \tau(1+\gamma)(1+\norm{z}_2).
\end{equation}
\end{lemma}
\begin{proof}
Set $V = \hat{X} - X_0$. We want to bound $\norm{V}_F \leq \norm{\mathcal{P}_\mathcal{T} (V)}_{F} + \norm{\mathcal{P}_{\mathcal{T}^\perp} (V)}_{F}$. Since $\hat{X}$ is the minimizer of \eqref{eq:suf_cond_min_prob} we have
\begin{equation} \label{eq:A(V)_2}
    \norm{\mathcal{A}(V)}_2 \leq
    \norm{\mathcal{A}(\hat{X})-\hat{y}}_2 + \norm{\hat{y} - \mathcal{A}(X_0)}_2 \leq
    2\tau.
\end{equation}
Combined with the local isometry property \eqref{LIP_T}, $\gamma$ being the operator norm of $\mathcal{A}$ and the triangle inequality we get
\begin{equation}\label{eq:P_T_V_upper_bound}
    \begin{split}
        \norm{\mathcal{P}_\mathcal{T} (V)}_{F} &\leq
        \frac{1}{\sqrt{1-\delta}} \norm{\mathcal{A}(\mathcal{P}_\mathcal{T} (V))}_2  \\
        & \leq \frac{1}{\sqrt{1-\delta}} (\norm{\mathcal{A}(\mathcal{P}_{\mathcal{T}^\perp} (V))}_2 + \norm{\mathcal{A}(V)}_2) \\
        & \leq \frac{1}{\sqrt{1-\delta}} (\gamma \norm{\mathcal{P}_{\mathcal{T}^\perp} (V))}_2 + 2 \tau).
    \end{split}
\end{equation}
To upper bound $\norm{\mathcal{P}_{\mathcal{T}^\perp} (V)}_2$, we choose $Z_0 \in \mathcal{T}^\perp$ such that $\norm{Z_0} \leq 1-\beta$ and $\langle Z_0, V \rangle _F = (1-\beta) \norm{\mathcal{P}_{\mathcal{T}^\perp} (V)}_*$. This is possible due to the duality of the norms $\norm{\cdot}_{2\to 2}$ and $\norm{\cdot}_*$. Note that $\norm{ \sgn(X_0) + \mathcal{P}_{\mathcal{T}^\perp} (Y) + Z_0} \leq 1$ since $\sgn(X_0) \perp \mathcal{P}_{\mathcal{T}^\perp} (Y) + Z_0$, the mentioned bound on $\norm{Z_0}$, \eqref{eq:beta} and $\norm{\sgn(X_0)}\leq 1$.
% Notice that $\norm{X_0 + \mathcal{P}_{\mathcal{T}^\perp}(Y) + Z} \leq 1$
Using this duality again, we get
\begin{flalign}\label{eq:X_0_V_nuc_norm_step_1}
        &\norm{X_0 + V}_*  = \underset{Z\in \mathbb{C}^{K \times N}, \norm{Z}\leq1}{\sup} | \langle Z , X_0 + V \rangle_F|  \\ \nonumber
        & \geq Re (\langle \sgn(X_0) + \mathcal{P}_{\mathcal{T}^\perp} (Y) + Z_0 , X_0 + V \rangle_F )\\ \nonumber
        &  =  \norm{X_0}_* + Re(\langle \mathcal{P}_{\mathcal{T}^\perp} (Y) + Z_0  , X_0 \rangle_F)  +\\ \nonumber
        &    Re (\langle \sgn(X_0) + \mathcal{P}_{\mathcal{T}^\perp} (Y) ,  V \rangle_F ) + Re (\langle Z_0 ,  V \rangle_F )  \\ \nonumber
        &  = \norm{X_0}_*  +    Re (\langle \sgn(X_0) + \mathcal{P}_{\mathcal{T}^\perp} (Y) ,  V \rangle_F ) +\\ \nonumber
        &(1-\beta) \norm{\mathcal{P}_{\mathcal{T}^\perp} (Y)}_*  
            =  \norm{X_0}_* +(1-\beta) \norm{\mathcal{P}_{\mathcal{T}^\perp} (V)}_* +\\\nonumber
        &Re (\langle \sgn(X_0) - \mathcal{P}_{\mathcal{T}} (Y) ,  V \rangle_F + \langle Y ,  V \rangle_F), \nonumber
\end{flalign}
where the first equality is due to $\norm{X}_* = \langle \sgn(X), X \rangle_F$ (see \eqref{def:sgn_func}) and the third equality follows $Re(\langle \mathcal{P}_{\mathcal{T}^\perp} (V) + Z_0  , X_0 \rangle_F) = 0$. Notice, that our definition of the sign function in \eqref{def:sgn_func}, which differs from the one in \cite{linear_blind_deconv_demix_one_measurement}, is essential for this step.
The last step is due to $Y = \mathcal{P}_\mathcal{T} (Y) + \mathcal{P}_{\mathcal{T}^\perp} (Y)$.

We now examine the term $Re (\langle \sgn (X_0) - \mathcal{P}_{\mathcal{T}} (Y) ,  V \rangle_F)$ in the last line of \eqref{eq:X_0_V_nuc_norm_step_1}. By Cauchy-Schwarz, the upper bound for $\norm{\mathcal{P}_{\mathcal{T}} (V)}_F$ in \eqref{eq:P_T_V_upper_bound} and the assumption in \eqref{eq:alpha}, we get
\begin{equation}\label{eq:Re_bound_1}
    \begin{split}
         &Re (\langle \sgn(X_0) - \mathcal{P}_{\mathcal{T}} (Y) ,  V \rangle_F)  \\
         &\geq - \norm{\sgn(X_0) - \mathcal{P}_{\mathcal{T}} (Y)}_F \norm{\mathcal{P}_{\mathcal{T}} (V)}_F  \\
         & \geq\frac{-\alpha}{\sqrt{1-\delta}}(\gamma \norm{\mathcal{P}_{\mathcal{T}^\perp} (V))}_2 + 2 \tau)
    \end{split}
\end{equation}
where in the first inequality we have also used the fact that $\sgn(X_0) - \mathcal{P}_{\mathcal{T}} (Y) \in \mathcal{T}$.
To bound the term $Re (\langle Y, V \rangle_F)$ in the last line of \eqref{eq:X_0_V_nuc_norm_step_1}, note that by Cauchy-Schwarz and \eqref{eq:A(V)_2}, 
\begin{equation}\label{eq:Re_bound_2}
    \begin{split}
        Re(\langle Y, V \rangle_F) = 
        &Re(\langle \mathcal{A}^*(z), V \rangle_F) \\
        =& Re(\langle z, \mathcal{A}(V) \rangle_2) 
        \geq -2\norm{z}_2 \tau
    \end{split}
\end{equation}
Putting \eqref{eq:Re_bound_1} and \eqref{eq:Re_bound_2} back into \eqref{eq:X_0_V_nuc_norm_step_1}, we get
\begin{equation}
    \begin{split}
        \hat{\norm{X}}_* =& \norm{X_0 + V}_* \\
        \geq & \norm{X_0}_* +
              \bigg{(} 1- \beta - \frac{\alpha \gamma}{\sqrt{1-\delta}} \bigg{)} \norm{\mathcal{P}_{\mathcal{T}^\perp} (V)}_* \\
        -& 2\tau  \bigg{(} \norm{z}_2 + \frac{\alpha}{\sqrt{1-\delta}} \bigg{)}.
    \end{split}
\end{equation}
Since $\hat{X}$ is the minimizer of the Nuclear norm, we have $\hat{\norm{X}}_* \leq \norm{X_0}_*$ and therefore
\begin{eqnarray}
        \norm{\mathcal{P}_{\mathcal{T}^\perp} (V)}_* 
        \leq \frac{2\tau  \bigg{(} \norm{z}_2 + \frac{\alpha}{\sqrt{1-\delta}} \bigg{)}}
                    {\bigg{(} 1- \beta - \frac{\alpha \gamma}{\sqrt{1-\delta}} \bigg{)}}.
\end{eqnarray}
Considering our assumptions on the constants we get $\norm{\mathcal{P}_{\mathcal{T}^\perp} (V)}_F \lesssim \tau (\norm{z}_2 + 1)$.
% \begin{equation}
%         \norm{\mathcal{P}_{\mathcal{T}^\perp} (V)}_F \lesssim \tau (\norm{z}_2 + 1).
% \end{equation}
Finally, we can bound
\begin{eqnarray}
        &\norm{V}_F \leq \norm{\mathcal{P}_\mathcal{T} (V)}_{F} + \norm{\mathcal{P}_{\mathcal{T}^\perp} (V)}_{F} \\ \nonumber
        & \lesssim (1 + \gamma) \norm{\mathcal{P}_{\mathcal{T}^\perp} (V)}_{F} + \tau 
         \lesssim \tau (1 + \gamma) (\norm{z}_2 + 1)
\end{eqnarray}
\end{proof}
The error bound requires us to bound also $\gamma$, the operator norm of the measurement operator $\mathcal{A}$. This is done in the following lemma, with its proof in App. \ref{proof:operator_norm_bound}. 
\begin{lemma}[Operator norm bound]\label{lemma:operator_norm_bound}
Let $\omega \geq 1$. Then with probability of at least $1-2L^{-\omega}$,
\begin{equation}
   \norm{\mathcal{A}} \leq 2\max \bigg{\{} 1, \sqrt{\frac{NK}{L}}\mu \bigg{\}} \sqrt{\log (L+SKN)}
\end{equation}
\end{lemma}

\subsection{Local Isometry Property}
We now show that the measurement operators $\mathcal{A}, \mathcal{A}^p$ (\eqref{A}, \eqref{A_p}) act as approximate isometries on $\mathcal{T}, \mathcal{T}^p$ (Def. \ref{solution_space}, \ref{extended_solution_space}).
\begin{theorem}\label{LIP}
Fix $\omega \geq 1$. Suppose that 
\begin{equation}\label{eq:LIP_Q_condition}
    Q \geq C_\omega \delta^{-2}S\left(K \mu^2 \log(L)\log^2(K\mu^2) + N\mu_H^2\right),
\end{equation}
then with probability $1-\mathcal{O}(L^{-\omega})$ the operator $\mathcal{A}$ satisfies \eqref{LIP_T} (LIP),
% \begin{equation}\label{LIP_T}
%     (1-\delta)\norm{X}_F^2 \leq \norm{\mathcal{A}(X)}_2^2 \leq (1+\delta)\norm{X}_F^2
% \end{equation}
and for all $p \in [P]$, every $Y \in \mathcal{T}^p = \mathcal{T} +\mathcal{T}_{S^pH}$ fulfills
\begin{equation}\label{LIP_p}
    (1-\delta)\norm{T_p^{\frac{1}{2}}Y}_F^2 \leq \frac{L}{Q} \norm{\mathcal{A}^p(Y)}_2^2 \leq (1+\delta)\norm{T_p^{\frac{1}{2}}Y}_F^2,
\end{equation}
where $T_p^{1/2}$ denotes the unique positive, self-adjoint matrix whose square is equal to $T_p$.
\end{theorem}
To prove this we need to define the following norms.
\begin{definition}\label{B_norms}
For any vector $z \in \mathbb{C}^K$ and matrix $Z \in \mathbb{C}^{K\times N}$:
\begin{equation}\label{def:B_norm}
    \norm{z}_B = \sqrt{L} \underset{l\in[L]}{\max}|z^*b_l|, \norm{Z}_B = \sqrt{L} \underset{l\in[L]}{\max} \norm{Z^*b_l}_2.
\end{equation}
% \[ \norm{z}_B = \sqrt{L} \underset{l\in[L]}{\max}|z^*b_l| \],
% \[ \norm{Z}_B = \sqrt{L} \underset{l\in[L]}{\max} \norm{Z^*b_l}_2. \]
\end{definition}
Our strategy is to use the following proposition, proven in App. \ref{proof:suprema_of_chaos_modification} and based on Th.~\ref{theorem:suprema_of_chaos_processes} regarding suprema of chaos processes. This involves the $\gamma_2$ functional, a geometric quantity introduced by Talagrand \cite{talagrand} and defined here in Def.~\ref{def:gamma_2}, and the distance $d_B(\mathcal{Z}) = \underset{Z\in \mathcal{Z}}{\sup} \norm{Z}_B$ (similarly for the Frobenius norm). This is further discussed in App. \ref{sec:sup_of_chaos_covering_numbers}.

\begin{proposition}\label{suprema_of_chaos_modification}
    Let $\mathcal{Z} \subset \mathcal{M}$ be a symmetric set and 
    \begin{gather*}
        E = \frac{\gamma_2 (\mathcal{Z}, \norm{\ \cdot \ }_B)}{\sqrt{Q}} \bigg{(} \frac{\gamma_2 (\mathcal{Z}, \norm{\ \cdot \ }_B)}{\sqrt{Q}} + d_F(\mathcal{Z}) \bigg{)} \\
        V = \frac{d_B(\mathcal{Z})}{\sqrt{Q}} \bigg{(} \frac{\gamma_2 (\mathcal{Z}, \norm{\ \cdot \ }_B)}{\sqrt{Q}} + d_F(\mathcal{Z}) \bigg{)} \\
        U = \frac{1}{Q} d_B^2(\mathcal{Z})
    \end{gather*}
    Then for $t \geq 0$ and all $p \in P$,
    \begin{equation}\label{sup_p}
        \begin{split}
        \mathbb{P} \bigg{(} \underset{Z \in \mathcal{Z}}{\sup} \bigg{|} \frac{L}{Q} \norm{\mathcal{A}^p(Z)}_2^2 - \norm{T_P^{1/2}Z}_F^2 \bigg{|} \geq c_1E+t \bigg{)} \leq \\ 2\text{exp}\bigg{(} -c_2 \ \min \bigg{(} \frac{t^2}{V^2}, \frac{t}{U} \bigg{)} \bigg{)}
        \end{split}
    \end{equation}
    \begin{equation}\label{sup_L}
        \begin{split}
        \mathbb{P} \bigg{(} \underset{Z \in \mathcal{Z}}{\sup} \bigg{|} \frac{L}{Q} \norm{\mathcal{A}(Z)}_2^2 - \norm{Z}_F^2 \bigg{|} \geq c_3E+t \bigg{)} \leq \\ 2\text{exp}\bigg{(} -c_4 \ \min \bigg{(} \frac{t^2}{V^2}, \frac{t}{U} \bigg{)} \bigg{)}
    \end{split}
    \end{equation}
    given that $\{ \Gamma_P \}_{p \in P}$ is an $\omega$-admissible partition of $[L]$. 
\end{proposition}

We now apply the proposition on the appropriate sets for proving Th. \ref{LIP}. We define the solution subspaces
\begin{gather}\label{eq:ball_definitions}
    \mathcal{B}^M = \big{\{} X \in \mathcal{T}_M \ |\ \norm{X}_F \leq 1 \big{\}} \\ \nonumber
    \mathcal{B}^H = \big{\{} X \in \mathcal{T}_H \ |\ \norm{X}_F \leq 1 \big{\}} \\ \nonumber
    \mathcal{B}^{S^pH} = \big{\{} X \in \mathcal{T}_{S^pH} \ |\ \norm{X}_F \leq 1 \big{\}}.
\end{gather}
Note, that these are sets of rank-$S$ matrices. The LIP in Th. \ref{LIP} follows by applying Proposition~\ref{suprema_of_chaos_modification} on the set (in place of $\mathcal{Z}$)
\begin{equation}
    \mathcal{W} =  \mathcal{B}^M + \mathcal{B}^H 
\end{equation}
and in a similar way we get \eqref{LIP_p} by applying it on the set
\begin{equation}
    \mathcal{W}^p =  \mathcal{W} +\mathcal{B}^{S^pH}.
\end{equation}
Thus, we only need to estimate the $\gamma_2$-functional, $d_B(\mathcal{Z})$ and $d_F(\mathcal{Z})$ on these sets, which is provided by the following lemma, with its proof in App. \ref{proof:suprema_of_chaos_bounds}.

\begin{lemma}\label{lemma:suprema_of_chaos_bounds}
Suppose that $\mathcal{X} = \mathcal{W}$ or  $\mathcal{X} = \mathcal{W}^p$ for some $p \in [P]$. Then 
\begin{gather}
        d_F(\mathcal{X}) \leq 3, \label{eq:ineq_1}\\
        d_B(\mathcal{X}) \leq 3 \sqrt{K}\mu, \label{eq:ineq_2} \\
        \gamma_2(\mathcal{X}, \norm{\cdot}_B) \lesssim \sqrt{S(K \mu^2 \ \log(L)\log^2(K \mu^2) + N \mu_H^2)} \label{eq:ineq_3}.
\end{gather}
% where $ d_B(\mathcal{X}) = \underset{X \in \mathcal{X}}{sup} \norm{X}_B$ (similarly for $d_F(\mathcal{X})$). 
\end{lemma}

Now we can prove Th. \ref{LIP}.
\begin{proof}[Proof of Th.~\ref{LIP}] 
Fix $p\in [P]$. Using Lemma \ref{lemma:suprema_of_chaos_bounds} 
and choosing the constant $C_\omega$ in \eqref{eq:LIP_Q_condition} large enough we get $E\leq \frac{\delta}{2c_1}$, $V\leq \frac{\delta}{\sqrt{c_2\omega \log L}}$ and $U\leq \frac{\delta}{c_2\omega \log L}$, where $\mathcal{X} \subset \mathcal{W}^p$. The inequality \eqref{sup_p} in Proposition \ref{suprema_of_chaos_modification} with $t=\frac{\delta}{2}$ shows that \eqref{LIP_p} in Th. \ref{LIP} holds with probability of $1-\mathcal{O}(L^{-\omega})$ (same holds for \eqref{sup_L} and \eqref{LIP_T}, with $\mathcal{X} \subset \mathcal{W}$). Replacing $\omega$ by $\omega +1$ and using a union bound argument shows that \eqref{LIP_p} and \eqref{LIP_T} are satisfied for all $p\in[P]$ with a probability of at least $1-(P+1)\mathcal{O}(L^{-\omega-1}) = 1-\frac{P+1}{L}\mathcal{O}(L^{-\omega})= 1- \mathcal{O}(L^{-\omega})$, which finishes the proof.
\end{proof}

\subsection{Constructing the Dual Certificate.}
%\label{sec:dual_cert}
We now turn to prove that the assumptions in Lemma \ref{sufficient_condition_for_recovery} hold.
As in previous works, we construct the dual certificate via the Golfing Scheme, presented in \cite{Golfing_Scheme}. Thus, we build the dual certificate with the iterative process
\begin{gather*}
    Y_0 = 0,\\ 
    Y_p = Y_{p-1} + \frac{L}{Q} (\mathcal{A}^p)^* \mathcal{A}^p S^p (\sgn(X_0) - \mathcal{P}_\mathcal{T}(Y_{p-1})),
\end{gather*}
where the final certificate $Y$ is given by
\begin{equation}
    Y = Y_P  = \sum_{p=1}^P \frac{L}{Q} (\mathcal{A}^p)^* \mathcal{A}^p S^p W_{p-1},
\end{equation}
and
\begin{equation}\label{def:W_p}
    W_p \triangleq \sgn(X_0) - \mathcal{P}_\mathcal{T}(Y_p).
\end{equation}

First, we show that $Y \in \range(\mathcal{A}^*)$. Recall that $\mathcal{A}^p$ is defined in \eqref{A_p} by taking only the measurements indexed by $l \in \Gamma_p$, while having zeros in the other entries. Thus, 
\begin{equation}
    (\mathcal{A}^p)^* \mathcal{A}^p S^p W_{p-1} = \mathcal{A}^* \mathcal{A}^p S^p W_{p-1}
\end{equation}
and we can write $Y$ as
\begin{equation}\label{dual_certificate}
    Y = \mathcal{A}^* \sum_{p=1}^P \frac{L}{Q} \mathcal{A}^p S^p W_{p-1},
\end{equation}
which is clearly in $\range(\mathcal{A}^*)$. It thus remains to show that assumptions \eqref{eq:alpha}, \eqref{eq:beta} in Lemma \ref{sufficient_condition_for_recovery} hold.

\subsubsection{Solution Space Bound}
We now show that condition \eqref{eq:alpha} holds. Despite the different definitions and rank, the proof in \cite{linear_blind_deconv_demix_one_measurement} is adequate also in our case, with very slight modifications.
We start with stating a private case of a technical lemma and continue to prove this section's main claim.
\begin{lemma}[a private case of Lemma 30 in \cite{linear_blind_deconv_demix_one_measurement}]
Let $\nu \leq \frac{1}{32}$. Then for all $p\in[p]$, 
\begin{equation}\label{eq:Tp^1/2}
    \norm{I - T_p^{1/2}} \leq\frac{1}{32}
\end{equation}
\begin{equation}\label{eq:I_S_p_norm}
    \norm{(I - S_p)X}_F \leq\frac{1}{31} \norm{X}_F
\end{equation}
\begin{equation}\label{eq:S_p_norm}
    \norm{S_p X}_F \leq\frac{32}{31} \norm{X}_F
\end{equation}
\end{lemma}
This allows us to prove the following lemma.

\begin{lemma}\label{lemma:solution_space_bound}
Suppose that $\mathcal{A}_p$ satisfies the $\delta$-local isometry property on $\mathcal{T}_p$ with $\delta=\frac{1}{32}$ for all $p\in[P]$. Then, for all $p\in[P]$,
\begin{equation}\label{eq:W_p_bound}
    \norm{W_p}_F \leq 4^{-p}\sqrt{S}
\end{equation}
and, in particular, if $P \geq \frac{1}{2} \log(8 \gamma \sqrt{S})$, 
\begin{equation}\label{eq:W_P_bound}
    \norm{\sgn(X_0) - \mathcal{P}_\mathcal{T}(Y)}_F \leq \frac{1}{8\gamma}.
\end{equation}
\end{lemma}
\begin{proof}
By \eqref{eq:T_p_I} and the triangle inequality we have 
\begin{equation*}
    (1-\nu)\norm{X}_F \leq \norm{T_p^{1/2} X}_F \leq (1+\nu)\norm{X}_F.
\end{equation*}
Combined with the $\delta$-local isometry property on $\mathcal{T}_p$ in \eqref{LIP_p},
\begin{equation*}
    (1-\nu)^2(1-\delta)\norm{X}_F^2 \leq \frac{L}{Q} \norm{\mathcal{A}^p(X)}_F \leq (1+\nu)^2(1+\delta)\norm{X}_F^2
\end{equation*}
for all $X \in \mathcal{T}^p$. With $\delta = \nu = \frac{1}{32}$, this implies
\begin{equation*}
    \bigg{|}\frac{L}{Q} \norm{\mathcal{A}^p(X)}_2^2 - \norm{X}_F^2 \bigg{|} \leq \frac{1}{8} \norm{X}_F^2
\end{equation*}
for all $X \in \mathcal{T}^p$, which is equivalent to 
\begin{equation}\label{eq:1_8}
    \norm{\mathcal{P}_{\mathcal{T}^p} - \frac{L}{Q} \mathcal{P}_{\mathcal{T}^p} (\mathcal{A}^p)^* \mathcal{A}^p \mathcal{P}_{\mathcal{T}^p} } \leq \frac{1}{8}.
\end{equation}

Notice, that by its definition in \eqref{def:W_p}, we have that
\begin{equation}\label{W_p_rec}
    W_p = W_{p-1} - \frac{L}{Q} \mathcal{P}_\mathcal{T} (\mathcal{A}^p)^* \mathcal{A}^p S^p W_{p-1}
\end{equation}
and also that $\norm{W_{p-1} - \mathcal{P}_\mathcal{T}(X)}_F \leq \norm{W_{p-1} - \mathcal{P}_{\mathcal{T}^p}(X)}_F $ for all $X \in \mathcal{M}$ since $W_{p-1} \in \mathcal{T}$ and $\mathcal{T} \subset \mathcal{T}^p$. This implies that
\begin{equation}
    \begin{split}
        \norm{W_p}_F &\leq \norm{W_{p-1} - \frac{L}{Q} \mathcal{P}_{\mathcal{T}^p} (\mathcal{A}^p)^*                            \mathcal{A}^p S^p W_{p-1}}_F \\
                     &= \norm{W_{p-1} - \frac{L}{Q} \mathcal{P}_{\mathcal{T}^p} (\mathcal{A}^p)^*                            \mathcal{A}^p \mathcal{P}_{\mathcal{T}^p} S^p W_{p-1}}_F,
    \end{split}
\end{equation}
where the equality is due to $S^p W_{p-1} \in \mathcal{T}_p$ and $W_{p-1} \in \mathcal{T}$. Combining this with \eqref{eq:I_S_p_norm}, \eqref{eq:S_p_norm} and \eqref{eq:1_8}, leads to 
\begin{flalign}
        &\norm{W_p}_F \leq \norm{ \bigg{(} I - \frac{L}{Q} \mathcal{P}_{\mathcal{T}^p} (\mathcal{A}^p)^*                            \mathcal{A}^p\bigg{)} S^p W_{p-1}}_F \\ \nonumber
                     &+\norm{(I - S^p) W_{p-1}}_F 
                     \leq \frac{1}{8} \norm{S^p W_{p-1}}_F +
                           \frac{1}{16} \norm{W_{p-1}}_F \\ \nonumber
                     & \leq \frac{1}{4} \norm{W_{p-1}}_F.
\end{flalign}
Thus, $\forall p\in[P]$, $\norm{W_p}_F \leq (1/4 )^p \norm{W_0}_F = (1/4)^p \sqrt{S}$
% \begin{equation}
%     \norm{W_p}_F \leq \bigg{(} \frac{1}{4} \bigg{)}^p \norm{W_0}_F = \bigg{(} \frac{1}{4} \bigg{)}^p \sqrt{S},
% \end{equation}
which proves \eqref{eq:W_p_bound}. 
% Notice, that $\norm{W_0}_F = \norm{\sgn(X_0)}_F = \sqrt{S}$ remains true despite the different sign definition in \cite{linear_blind_deconv_demix_one_measurement} (see \eqref{def:sgn_func}). 
As $P \geq \frac{1}{2}\log (8\gamma \sqrt{S})$ (Def. \ref{w_admissible}), we get \eqref{eq:W_P_bound}.
\end{proof}
\subsubsection{Outer Space Bound}
We now turn to show that condition \eqref{eq:beta} in Lemma~\ref{sufficient_condition_for_recovery} holds. Thus, we bound the operator norm:
\begin{equation*}
\begin{split}
    \norm{\mathcal{P}_{\mathcal{T}^\perp}(Y_P)} & \leq 
    \sum_{p=1}^P \norm{\mathcal{P}_{\mathcal{T}^\perp} \bigg{(} \frac{L}{Q} (\mathcal{A}^p)^* \mathcal{A}^p S^p W_{p-1} - W_{p-1} \bigg{)}}\\ &\leq
    \sum_{p=1}^P \norm{ \frac{L}{Q} (\mathcal{A}^p)^* \mathcal{A}^p S^p W_{p-1} - W_{p-1}},
\end{split}
\end{equation*}
where we use $W_{p-1} \in \mathcal{T}$ and the fact that the operator norm of a projection is bounded by $1$. Thus, in order to show that condition \eqref{eq:beta} holds, it remains to prove that
\begin{equation}\label{eq:outer_space_norm_to_bound}
    \norm{  \frac{L}{Q} (\mathcal{A}^p)^* \mathcal{A}^p S^p W_{p-1} - W_{p-1}}
     \leq \frac{1}{4^{p+1}},
\end{equation}
for $p\in[P]$. By defining
\begin{equation}\label{mu_p}
    \mu_p = \sqrt{L} \underset{l\in\gamma_p}{\max} \norm{W^*_p S_{p+1} b_l}_2,
\end{equation}
we state the outer space bound lemma (proven in App. \ref{proof:outer_space_bound}).
\begin{lemma}\label{lemma:outer_space_bound}
Let $\omega \geq 1$. Assume that
$\mu_p \leq 4^{-p} \mu_H$ and $\norm{W_p}_F \leq 4^{-p}\sqrt{S}$. If 
\begin{equation}
    Q \gtrsim S (K\mu^2 + N\mu_H^2)\log^2 L,
\end{equation}
then with probability $1-\mathcal{O}(L^{-\omega})$,
\begin{equation}
\norm{ \frac{L}{Q} (\mathcal{A}^p)^* \mathcal{A}^p S^p W_{p-1} - W_{p-1}} \leq \frac{1}{4^{p+1}}, ~~~~ \forall p\in[P].
\end{equation}
\end{lemma}

\subsubsection{An Upper Bound for the Dual Certificate}
The upper bound in \eqref{X_err_bound} scales with $\norm{z}_2$, where $z$ equals
\begin{equation}\label{z}
    z = \frac{L}{Q} \sum_{p=1}^P \mathcal{A}^p S^p W_{p-1},
\end{equation}
such that $Y = \mathcal{A}^* z$ as in \eqref{dual_certificate}. We thus need to upper bound $\norm{z}_2$ to obtain the total error bound.
\begin{lemma}\label{z_bound}
Let $z\in\mathcal{C}^L$ be given by \eqref{z} and assume $\norm{W_p}_F \leq 4^{-p} \sqrt{S}$. Suppose that $\mathcal{A}^p$ satisfies \eqref{LIP_p} with $\delta\leq \frac{1}{4}$ on $\mathcal{T}^p$ for all $p\in[P]$. Then $\norm{z}_2 \lesssim P \sqrt{S}$.
% \begin{equation}
%     \norm{z}_2 \lesssim P \sqrt{S}.
% \end{equation}
\end{lemma}
\begin{proof}
By its definition \eqref{z}, 
\begin{eqnarray*}
    \norm{z}_2 =  \frac{L}{Q} \sum_{p=1}^P \norm{ \mathcal{A}^p S^p W_{p-1}}_F
            \lesssim P \sum_{p=1}^P \norm{W_{p-1}}_F 
            % &\leq P \sum_{p=1}^P 4^{-p} \sqrt{S} 
            \lesssim P \sqrt{S},
            % \leq \log(8\gamma \sqrt{S}) \sqrt{S}\\
            % &\lesssim \log (L) \sqrt{S},
\end{eqnarray*}
where we have used \eqref{LIP_p}, \eqref{eq:S_p_norm} and $P=L/Q$ (Def. \ref{w_admissible}).
\end{proof}
Now we can finally prove the main Theorem. 

\subsection{Proof of Th. \ref{main_theorem}}
Combining the conditions on $Q$ given in Th. \ref{LIP}, Lemmas \ref{lemma:outer_space_bound} and \ref{lemma:mu_p}, we have that
\begin{equation}\label{total_Q_cond}
     Q \gtrsim S \left(K \mu^2 \log(K\mu^2) + N\mu_H^2\right) \log^2(L).
\end{equation} 
Let $\Gamma_p$ be an admissible partition of the measurements and let $\omega>0$. Then by Def. \ref{w_admissible} we have
\begin{equation}
    P \leq \log(8\gamma \sqrt{S}) \lesssim \log L,
\end{equation}
where $\gamma \leq 2\max \big{\{} 1, \sqrt{\frac{NK}{L}}\mu \big{\}} \sqrt{\omega(\log (L+SKN))}$. As $L=PQ$, we have that if
\begin{equation}
    L \gtrsim_\omega S \left(K \mu^2 \log(K\mu^2) + N\mu_H^2\right) \log^3(L)
\end{equation}
then we can assume that Th. \ref{LIP} and Lemmas \ref{lemma:outer_space_bound} and \ref{lemma:mu_p} hold.
Thus, we can assume that conditions \eqref{LIP_T} and \eqref{LIP_p} hold with prob. $1-\mathcal{O}(L^{-\omega})$ and constant $\delta = 1/32$. By applying Lemma \ref{sufficient_condition_for_recovery} with $\alpha=1/8\gamma$, $\beta=1/4$ and $\delta=1/4$, it is enough to construct a dual certificate $Y\in \range{\mathcal{A}^*}$, which satisfies conditions \eqref{eq:alpha} and \eqref{eq:beta}. These conditions are met by the Golfing scheme in Lemmas \ref{lemma:solution_space_bound} and \ref{lemma:outer_space_bound} for a fixed $p\in[P]$. The assumptions of Lemma \ref{lemma:solution_space_bound} are given by \eqref{LIP_p}, thus, condition \eqref{eq:alpha} applies. The assumptions of Lemma \ref{lemma:outer_space_bound} are met by Lemma \ref{lemma:mu_p}, which holds by \eqref{total_Q_cond} and so condition \eqref{eq:beta} holds. 
Thus, $Y$ defined in \eqref{dual_certificate} satisfies conditions \eqref{eq:alpha} and \eqref{eq:beta}. Using a union bound we conclude that with probability $1-\mathcal{O}(L^{-\omega})$ the approximate dual certificate satisfies the conditions of Lemma \ref{sufficient_condition_for_recovery} and thus if $\hat{X}$ is the minimizer of \eqref{eq:suf_cond_min_prob} then we can bound the estimation error by \eqref{X_err_bound}. This error is bounded by lemmas \ref{lemma:operator_norm_bound} and \ref{z_bound}, resulting in 
\begin{equation}
\begin{split}
    \norm{\hat{X} - X_0}_F &\lesssim \tau(1+\gamma)(1+\norm{z}_2) \\
                           &\lesssim_\omega \tau \sqrt{S \max \bigg{\{}1, \frac{SK\mu^2N}{L}\bigg{\}} \log^3 (L)}
\end{split}
\end{equation}

\section{Experiments}\label{sec:experiments}
\begin{figure}
      \centering
      \centerline{\includegraphics[width=0.6\linewidth]{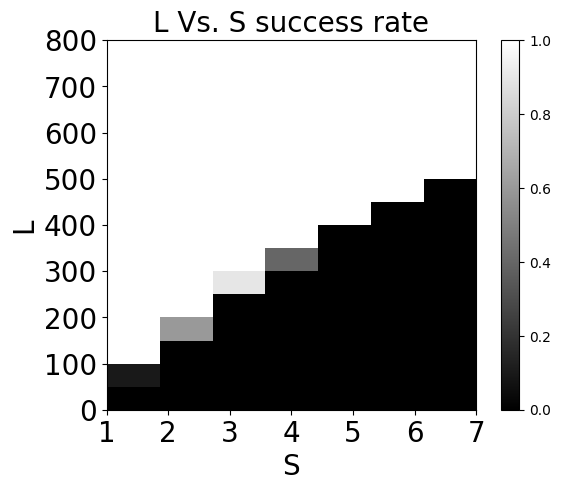}}
      \caption{Phase transition: linear empirical dependence of L in S, in accordance with our theoretical results.}
      \label{fig:L_S}
\end{figure}

\begin{figure}
    \centering
    \centerline{\includegraphics[width=0.6\linewidth]{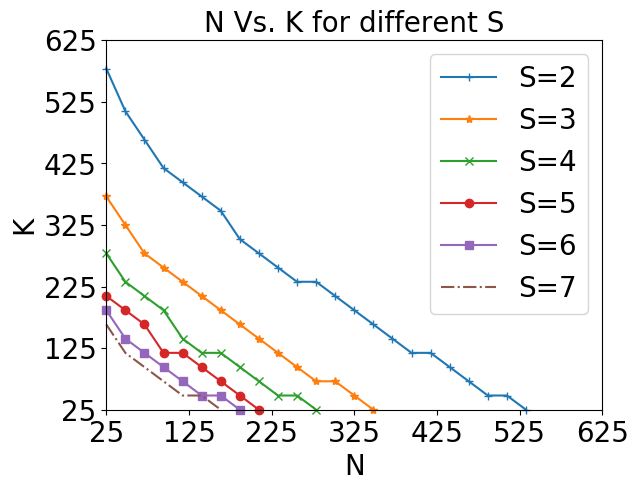}}
    \caption{Phase transition limits for a fixed $L$, varying the subspace dimensions N and K. The cutoffs match the theory.}
    \label{fig:N_K}
\end{figure}

In all the experiments, $C, M$, and $H_r$ are drawn from a random Gaussian distribution. $B$ consists of the $K$ first standard basis vectors, representing the blind MIMO scenario in \cite{blind_mimo_AA} and in accordance with the results presented in \cite{blind_deconv_demix}. The matrix reconstruction phase is done using the Matlab solver minfunc, using the heuristic solver developed by Burer and Monteiro \cite{Burer_Monteiro} (similar to \cite{subspace_prior_blind_deconv}). For the vector recovery, we used the Matlab non-linear equation solver fsolve with the trust-region Dogleg algorithm. All the results are measured end to end, ie. with an average error of less than $0.1\%$ for each vector. We repeat each experiment and report the success fraction out of 10 runs.

First, we show the phase transition of the empirical reconstruction probability, changing the number of measurements at each receiver, $L$, for a different number of signals. We fix $N=30, K=25$, and report the fraction of successful reconstructions (i.e. with an average error of less than $0.1\%$ for each vector) out of 10 experiments. The results for the two-sided constraints scenario, where $R=S$,  are shown in Fig. \ref{fig:L_S}, demonstrating linear dependency in accordance with the theoretical guarantees. These results are very similar to the phase transition empirical results presented in \cite{blind_deconv_demix} but require less computational resources to achieve them.

Next, we repeat a similar experiment, only this time fixing $L=2048$ and changing $N$ and $K$. We consider the two-sided constraints case, with $R=S$.  Fig. \ref{fig:N_K} shows phase transition lines given the same success criterion as before, for a different number of sources. The area below the line indicates successful recovery and shows the maximal $N, K$ for a given amount of measurements per receiver. For $S=2$, $L\sim4(N+K)$. For a larger $S=7$, $L\sim11(N+K)$. The "cutoff" $N, K$ appears to be nonlinear w.r.t. $S$ and differs more for smaller values of $S$. This result is in accordance with our linear theoretical guarantees.

\begin{figure}
\centering
\subfloat{\includegraphics[width=0.46\linewidth]{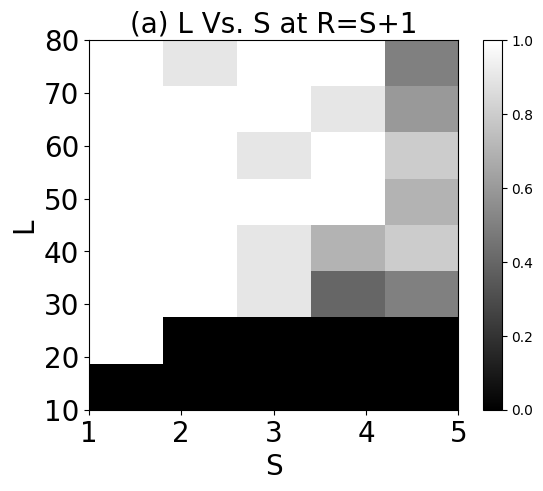}
\label{fig:R_S_p1}}
\hfil
\subfloat{\includegraphics[width=0.46\linewidth]{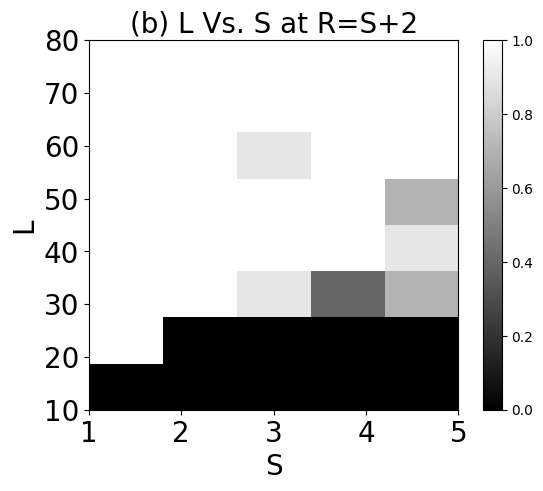}
\label{fig:R_S_p2}}
\vfil
\subfloat{\includegraphics[width=0.46\linewidth]{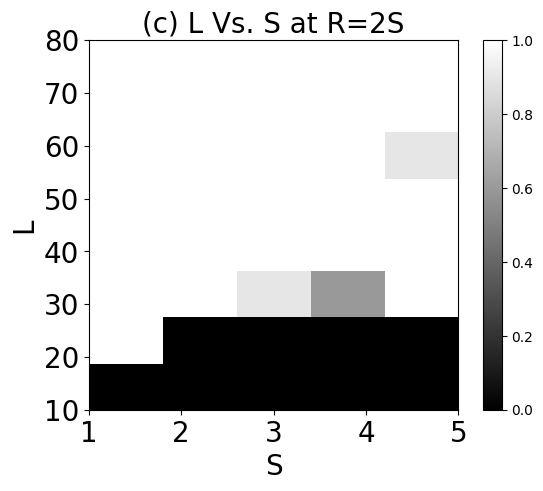}
\label{fig:R_2S}}
\caption{Phase transition for one sided constraints. More constraints improve the basis transformation optimization.}
\label{fig:one_sided_constraints}
\end{figure}

Fig. \ref{fig:one_sided_constraints} shows the phase transition results for the one sided constraints. We use $N=6, K=5, L=10,...,80$. Fig. \ref{fig:R_S_p1} presents the results for $R=S+1$, the minimal $R$ that allows correct recovery. The partial success areas (gray rubrics) in this chart are mostly failures to converge to some solution of the quadratic system \eqref{eq:one_sided_eq} (the vector recovery stage), as opposed to wrong ambiguous solutions. A different algorithm might suggest further improved performance. Figs. \ref{fig:R_S_p2} ($R=S+2$) and \ref{fig:R_2S} ($R=2S$) show that adding constraints, improves the convergence and precision.
The better performance of the two-sided setup (has fewer constraints compared to the ones sided case with $R> S+1$) might indicate that its structure does add up to more than the sum of its parts, and imposes some "hidden" constraints.

\section{Conclusion}
This work presents a separable approach to blind deconvolution and demixing via convex optimization. We measure the same signals at different receivers, with different convolution kernels. Assuming all signals and kernels reside in the same subspaces, allows formulating the problem as low-rank matrices recovery. Using the assumption that the signals are normalized allows us to recover them blindly.

Our formulation allows lower complexity than \cite{blind_mimo_AA} because we keep our matrices small. We solve a few small problems instead of a single large one. The stage we add to resolve the spanning base ambiguity of the rank-S case, is of a constant complexity of $S^2$, regardless of the number of receivers, and the other dimensions of the problem, i.e. $N, K, L$, which are usually much larger. Although we do not support the case of $R < S$, for the full case we have a much better complexity. We believe that our formulation can be combined with non-convex schemes\cite{Nonconvex_deconv_demix, Nonconvex_deconv_demix_2, nonconvex_hard_thresholding_quadratic, nonconvex_Riemannian_quadratic, nonconvex_quadratic_WF_randm_init, nonconvex_Wirtinger_Blair_quadratic} to further improve their complexity. We leave this to future work.

We derive sample complexity conditions for the matrix recovery problem given in \eqref{optimization_probelm}. This is the first work to solve this problem for rank-$S$ matrices and to supply adequate proof, which stands in line with the previous near-optimal results of the rank-$1$ case \cite{linear_blind_deconv_demix_one_measurement}.
Future work should analyze our conjecture resolving the spanning base ambiguity. We expect the bounds to remain linear, in line with our empirical results.

\appendices

\section{Suprema of Chaos Processes and Covering Numbers}\label{sec:sup_of_chaos_covering_numbers}
This section describes the necessary preliminaries used throughout the proof. We refer to \cite[Sec. IV-B,C]{linear_blind_deconv_demix_one_measurement} for further reading and references.
First, we define the $\gamma_2$ functional, a geometric quantity introduced by Talagrand \cite{talagrand}.
\begin{definition}\label{def:gamma_2}
Let $(X, ||| \cdot |||)$ be a Banach space and suppose that $S \subset X$.  A sequence $(S_n)_{n\geq0}$ of subsets of S is admissible, if $|S_0| = 1$ and $|S_n| = 2^{2^n}$ for $n\geq1$. Then we set 
\begin{equation}
    \gamma_2(S,  ||| \cdot |||) = \underset{(S_n)_{n\geq0}}{\inf} \underset{s\in S}{\sup} \sum_{n=0}^\infty 2^{n/2} \underset{s_n \in S_n}{\inf} ||| s - s_n|||,
\end{equation}
where the infimum is over all admissible sequences $(S_n)_{n\geq0}$.
\end{definition}

Furthermore, we define the distances
\begin{equation}
    d_F(\mathcal{X}) = \underset{X\in \mathcal{X}}{\sup} \norm{X}_F,
    d_{2\to2}(\mathcal{X}) = \underset{X\in \mathcal{X}}{\sup} \norm{X},
\end{equation}
where $\mathcal{X}$ is any set of matrices. Together with the notion of the $\gamma_2$ functional, we can state the following theorem that is used in the proof of Th. \ref{LIP}.

\begin{theorem}[Suprema of Chaos Processes; Th. 13 in \cite{linear_blind_deconv_demix_one_measurement}]\label{theorem:suprema_of_chaos_processes}
Let $\mathcal{X}$ be a symmetric set of matrices and let $\xi$ be a random vector whose entries are $\xi_i \sim \mathcal{CN}(0, 1)$ are independent. Set
    \begin{gather*}
        E = \gamma_2 (\mathcal{X}, \norm{\ \cdot \ })\big{(} \gamma_2 (\mathcal{X}, \norm{\ \cdot \ }) + d_F(\mathcal{X}) \big{)} \\
        V = d_{2\to2}(\mathcal{X}) \big{(} \gamma_2 (\mathcal{X}, \norm{\ \cdot \ }) + d_F(\mathcal{X}) \big{)} \\
        U = d^2_{2\to2}(\mathcal{X})
    \end{gather*}
    Then for $t \geq 0$,
    \begin{equation}
        \begin{split}
        \mathbb{P} \bigg{(} \underset{X \in \mathcal{X}}{\sup} \bigg{|} \norm{A\xi}_2^2 - \mathbb{E}\norm{A\xi}_2^2  \bigg{|} \geq c_1E+t \bigg{)} \leq \\ 2\text{exp}\bigg{(} -c_2 \ \min \bigg{(} \frac{t^2}{V^2}, \frac{t}{U} \bigg{)} \bigg{)}
        \end{split}
    \end{equation}
    where the constants $c_1, c_2$ are universal.
\end{theorem}

The $\gamma_2$ functional can be bounded using Dudley's inequality, involving covering numbers. Recall that the covering number $N (S, ||| \cdot |||, \epsilon)$ is the minimum number of $||| \cdot |||$-balls with radius $\epsilon$ to cover the set $S$.  
\begin{theorem}[Dudley's Inequality, see \cite{talagrand} Prop. 2.2.10, \cite{DUDLEY}]\label{theorem:Dudley's_ineq}
Given a set $S$ in a Banach space $(X,||| \cdot |||)$, we have that 
\begin{equation*}
    \gamma_2(S, ||| \cdot |||) \lesssim \int_0^{d_{||| \cdot |||} (S)} \sqrt{\log N (S, ||| \cdot |||, \epsilon) d\epsilon},
\end{equation*}
where $d_{||| \cdot |||}(S) = \underset{x \in S}{\sup} ||| x |||$.
\end{theorem}

We now prove Proposition \eqref{suprema_of_chaos_modification}, which is a modification to Th. \ref{theorem:suprema_of_chaos_processes} and used to prove Th. \ref{LIP}.
\begin{proof}[Proof of Proposition \ref{suprema_of_chaos_modification}]\label{proof:suprema_of_chaos_modification}
First we prove \eqref{sup_p}. Fix $p \in [P]$. For $Z \in \mathcal{Z}$, let $H_Z \in \mathbb{C}^{L \times |\Gamma_p| N}$ be a "block diagonal" matrix, where each block in it indexed by $l \in \Gamma_p$ is the row vector $\sqrt{\frac{L}{Q}}b_l^*Z \in \mathbb{C}^{1 \times N}$. Notice that
\begin{eqnarray*}
    \norm{H_Z}_F^2 &=& 
    \frac{L}{Q} \sum_{l \in \Gamma_p} \norm{Z^*b_l}^2_2 = 
    \text{Tr}(ZZ^*T_p) = 
    \norm{T^{1/2}_pZ}_F^2, \\
    \norm{H_Z} &=& \sqrt{\frac{L}{Q}} \underset{l \in \Gamma_p}{\max} \norm{b_l^*Z}_2 \leq 
    \frac{1}{\sqrt{Q}}\norm{Z}_B.
\end{eqnarray*}
Let $\xi^{(p)}$ be the concatenation of all $c_l$ where $l \in \Gamma_p$ and $c_l$ is the $l^{th}$ column of $C^T$. Then
\begin{equation}
    \begin{split}
        \frac{L}{Q} \norm{\mathcal{A}^p(Z)}_2^2 = &
        \frac{L}{Q} \sum_{l \in \Gamma_p} |\mathcal{A}^p(Z)(l)|^2 = \\
        &\frac{L}{Q} \sum_{l \in \Gamma_p} | b_l^*Zc_l |^2 = 
        \norm{H_Z \xi^{(p)}}^2_2.
    \end{split}
\end{equation}
Notice that 
\begin{equation}
    \mathbb{E}\bigg{[}\norm{H_Z \xi^{(p)}}_2^2\bigg{]} = \norm{T_p^{1/2}Z}_F^2 = 
    \norm{H_Z}_F^2,
\end{equation}
and thus
\begin{equation}
    \begin{split}
    &\underset{Z \in \mathcal{Z}}{\sup} \bigg{|} \norm{H_Z \xi^{(p)}}^2_2 - \mathbb{E}\bigg{[}\norm{H_Z \xi^{(p)}}_2^2\bigg{]} \bigg{|} =\\
     &\underset{Z \in \mathcal{Z}}{\sup} \bigg{|} \frac{L}{Q} \norm{\mathcal{A}^p(Z)}_2^2 - \norm{T_P^{1/2}Z}_F^2 \bigg{|}.
    \end{split}
\end{equation}
To get \eqref{sup_p} we just need to apply Th. \ref{theorem:suprema_of_chaos_processes}. 
The proof for \eqref{sup_L} is similar, where the "diagonal" elements in $H_Z$ are non zero for $l \in[L]$, and $T_p$ is replaced with $\sum_{l \in [L]}b_lb^*_l = I$.
\end{proof} 

\section{Proof of Different Lemmas}

\subsection{Proof of Lemma \ref{lemma:operator_norm_bound}}\label{proof:operator_norm_bound}
\begin{proof}[Proof of Lemma \ref{lemma:operator_norm_bound}]
Recall $\mathcal{A}(Z)=[\langle A_1 , Z \rangle ,\ldots, \langle A_L , Z \rangle]^T $ where $A_l =  \sqrt{L} B^* F^* diag(F_l) \bar{F} \bar{C} = \hat{b}_l \hat{c}_l^*$. In order to bound $\gamma$, we will estimate the norms of the expected operator norms of $\mathcal{A^*A}$ and $\mathcal{AA^*}$. Starting with the former,
\begin{equation}
    \begin{split}
        &\mathbb{E}[\mathcal{A^*A}(Z)]  = \sum_{l \in [L]} \mathbb{E}[\mathcal{A}(Z)(l) \hat{b}_l\hat{c}_l^*]= \\
                                      & \sum_{l \in [L]} \mathbb{E}[\hat{b}_l \hat{b}_l^* Z \hat{c}_l \hat{c}_l^*]
                                       = \sum_{l \in [L]} \hat{b}_l \hat{b}_l^* Z = Z,
    \end{split}
\end{equation}
meaning that $\mathbb{E}[\mathcal{A^*A}] = \mathbb{I}$. Moving to the latter,
\begin{equation}
    \begin{split}
        &\mathbb{E}[\mathcal{AA^*}y(l)]  = \mathbb{E}[\hat{b}_l^* \mathcal{A}^*(y) \hat{c}_l]
                                       = \sum_{l^` \in [L]} \mathbb{E}[\hat{b}_l^* \hat{b}_{l^`} y(l^`) \hat{c}_{l^`}^* \hat{c}_l]\\
                                      &= y(l) \sum_{l^` \in [L]} \mathbb{E}[\hat{b}_l^* \hat{b}_{l^`} \hat{c}_{l^`}^* \hat{c}_l]
                                       = y(l) N \hat{\norm{b_l}}_2^2.
    \end{split}
\end{equation}
Thus, $\mathbb{E}[\mathcal{AA^*}] = \text{diag}(N\hat{\norm{b_1}}_2^2, \dots, N\hat{\norm{b_L}}_2^2)$. Combined with the definition of $\mu$ in \eqref{B_incoherence}, we get $\norm{\mathbb{E}[\mathcal{AA^*}]} \leq \frac{NK\mu^2}{L}$. Thus,
\begin{eqnarray}
        \sigma^2  = \max \{ \norm{\mathbb{E}[\mathcal{A^*A}]}, \norm{\mathbb{E}[\mathcal{AA^*}]} \} 
                  \leq  \max \bigg{\{} 1, \frac{NK\mu^2}{L} \bigg{\}}.
\end{eqnarray}
Applying Corollary 10 in \cite{linear_blind_deconv_demix_one_measurement} with $t=\omega \log(L)$,
\begin{eqnarray}
    \norm{\mathcal{A}} \leq \max \bigg{\{} 1, \sqrt{\frac{NK}{L}}\mu \bigg{\}} \sqrt{(2\omega \log L + \log (L+SKN))}
\end{eqnarray}
with probability exceeding $1-2L^{-\omega}$.
\end{proof}

\subsection{Proof of Lemma \ref{lemma:outer_space_bound}}\label{proof:outer_space_bound}
We will use Th. 9 in \cite{linear_blind_deconv_demix_one_measurement} to bound \eqref{eq:outer_space_norm_to_bound}.
\begin{lemma}[Matrix Bernstein Inequality, Th. 9 in \cite{linear_blind_deconv_demix_one_measurement}]\label{Matrix_Bernstein_Inequality}
Let $\alpha\in[1,\infty)$ and let $X_1, \dots, X_n \in \mathbb{C}^{d_1\times d_2}$ be independent random matrices that satisfy $\mathbb{E}[X_i]=0$ for all $i\in[n]$. 
Set $R_{\psi_\alpha}= \underset{i\in[n]}{\max}\norm{\norm{X_i}}_{\psi_\alpha}$ and 
$\sigma^2 = \max \Bigg{\{} \norm{ \sum_{i=1}^n \mathbb{E} [X_i X_i^*]}, \norm{ \sum_{i=1}^n \mathbb{E} [X_i^* X_i]} \Bigg{\}}$. 
Set $Z=\sum_{i=1}^n X_i$. Then with probability at least $1-e^{-t}$, 
\begin{flalign}
    &\norm{Z} \lesssim  \max \bigg{\{} \sigma \sqrt{t+\log(d_1+d_2)},\\ \nonumber
    &R_{\psi_\alpha}\bigg{(}\log \bigg{(}1+ \frac{n R_{\psi_\alpha}^2}{\sigma^2}\bigg{)}\bigg{)}^{1/\alpha} (t+\log(d_1+d_2))  \bigg{\}}.
\end{flalign}
\end{lemma}
\begin{proof}[Proof of Lemma \ref{lemma:outer_space_bound}]
First, notice that for all $l \in \Gamma_p$,
\begin{eqnarray}
    (\mathcal{A}^p S^p W_{p-1}) (l) &=& b_l^* S^p W_{p-1} c_l,\\ \nonumber
    ((\mathcal{A}^p)^*\mathcal{A}^p S^p W_{p-1}) &=& \sum_{l\in\Gamma_p} b_l b_l^* S^p W_{p-1} c_l c_l^*.
\end{eqnarray}
Since by its definition, $S_p = (T_p)^{-1}$ (see \eqref{S_p}), we have that
\begin{equation}
    W_{p-1} = T_p S_p W_{p-1} = \frac{L}{Q} b_l b_l^* S_p W_{p-1}.
\end{equation}
For simplicity of notation, define 
\begin{equation}\label{w_l}
    w_l = W^*_{p-1} S_p b_l.
\end{equation}
Thus, we can write
\begin{equation}\label{W_p_before_proj}
    \begin{split}
        & \frac{L}{Q} (\mathcal{A}^p)^* \mathcal{A}^p S^p W_{p-1} - W_{p-1} = 
         \frac{L}{Q} \sum_{l\in\Gamma_p} b_l w_l^* c_l c_l^* - \\
         &\frac{L}{Q} \sum_{l\in\Gamma_p} b_l w_l^* =
         \frac{L}{Q} \sum_{l\in\Gamma_p} b_l w_l^* (c_l c_l^* - I) = \sum_{l\in\Gamma_p} Z_l,
    \end{split}
\end{equation}
where 
\begin{equation}\label{Z_l}
    Z_l \triangleq \frac{L}{Q} b_l w_l^* (c_l c_l^* - I).
\end{equation}
We now asses the the relevant components to use Lemma \ref{Matrix_Bernstein_Inequality}.  We start with the expectation values of $Z_l Z_l^*$ and $Z_l^*Z_l$.
\begin{equation}
\begin{split}
    &\mathbb{E} [Z_l Z_l^*] = \mathbb{E} [\frac{L^2}{Q^2} b_l w_l^* (c_l c_l^* - I)^2 w_l b_l^*] \\
                           &= \frac{L^2}{Q^2} b_l w_l^* \mathbb{E}[(c_l c_l^* - I)^2] w_l b_l^* 
                           = \frac{L^2}{Q^2} N \norm{w_l}_2^2 b_l b_l^*,
\end{split}
\end{equation}
since $\mathbb{E}[(c_l c_l^* - I)^2] = N I$ by Lemma 11 in \cite{subspace_prior_blind_deconv}.
\begin{flalign}
    &\mathbb{E} [Z_l^* Z_l] = \mathbb{E} [\frac{L^2}{Q^2} (c_l c_l^* - I) w_l b_l^* b_l w_l^* (c_l c_l^* - I) ] =\\ \nonumber
                           & \frac{L^2}{Q^2} \norm{b_l}_2^2 \mathbb{E}[(c_l c_l^* - I)w_l w_l^*(c_l c_l^* - I)] 
                           = \frac{L^2}{Q^2} \norm{b_l}_2^2 \norm{w_l}_2^2 I,
\end{flalign}
since $\mathbb{E}[(c_l c_l^* - I)w_l w_l^*(c_l c_l^* - I)] = \norm{w_l}_2^2 I$ by Lemma 12 in\cite{subspace_prior_blind_deconv}. Furthermore, we have that
\begin{equation}
\begin{split}
   \norm{ \sum_{l \in \Gamma_p} \mathbb{E} [Z_l Z_l^*]} &\leq 
        \frac{L^2N}{Q^2} \underset{l \in \Gamma_p}{\max}(\norm{w_l}_2^2) \norm{ \sum_{l \in \Gamma_p} b_l b_l^*} \\
        &\leq \frac{N}{Q} \mu^2_{p-1} \norm{T_p} 
         \lesssim\frac{16^{-p}N\mu_H^2}{Q},
\end{split}
\end{equation}
due to the lemma's assumptions and \eqref{eq:T_p_I}.
\begin{equation}
\begin{split}
   &\norm{ \sum_{l \in \Gamma_p} \mathbb{E} [Z_l^* Z_l]} \leq 
        \frac{L^2}{Q^2} \underset{l \in \Gamma_p}{\max}(\norm{b_l}_2^2) \norm{ \sum_{l \in \Gamma_p} \norm{w_l}_2^2} \\
        &\leq \frac{LK\mu^2}{Q^2} \sum_{l \in \Gamma_p}  \tr (W^*_{p-1} S_p b_l b_l^* S_p W_{p-1} )\\
        &= \frac{K\mu^2}{Q} \norm{S_p^{1/2}W_{p-1}}_F^2 
        \lesssim 16^{-p} \frac{SK\mu^2}{Q}.
\end{split}
\end{equation}
Thus, we have
\begin{eqnarray}
     \sigma^2 &=& \max \Bigg{\{} \norm{ \sum_{l \in \Gamma_p} \mathbb{E} [Z_l Z_l^*]}, \norm{ \sum_{l \in \Gamma_p} \mathbb{E} [Z_l^* Z_l]} \Bigg{\}} \\ \nonumber &
     \lesssim &  \frac{16^{-p}}{Q} \max (SK\mu^2, N\mu_H^2) \leq \frac{16^{-p}}{Q}  (SK\mu^2+ N\mu_H^2).
\end{eqnarray}
Now we estimate $R_{\psi_\alpha} = \max \norm{\norm{Z_l}}_{\psi_\alpha}$, where $\norm{\cdot}_{\psi_\alpha}$ is the Orlicz norm (Def. 7 in \cite{linear_blind_deconv_demix_one_measurement}).
\begin{eqnarray}
\hspace{-0.3in}     &&\norm{\norm{Z_l}}_{\psi_1} = \norm{\norm{\frac{L}{Q} b_l w_l^* (c_l c_l^* - I)}}_{\psi_1} \\ \nonumber
\hspace{-0.3in}     &&\leq \frac{L}{Q} \norm{\norm{ b_l w_l^* (c_l c_l^* - I)}_2}_{\psi_1} 
     \leq \frac{L}{Q} \norm{ b_l}_2 \norm{\norm{w_l^* (c_l c_l^* - I)}_2}_{\psi_1} \\ \nonumber
\hspace{-0.3in}     &&\lesssim \frac{L\sqrt{N}}{Q} \norm{ b_l}_2 \norm{ w_l}_2 
     \lesssim \frac{L\sqrt{N}}{Q} \sqrt{\frac{K}{L}} \mu \frac{\mu_{p-1}}{\sqrt{L}} \\ \nonumber
\hspace{-0.3in}     &&\lesssim 4^{-p} \frac{\sqrt{NK}\mu \mu_H}{Q} 
     \lesssim 4^{-p} \frac{(K\mu^2+ N\mu_H^2)}{Q},
\end{eqnarray}
where the second inequality is due to Lemma 39 in \cite{linear_blind_deconv_demix_one_measurement}, and the last step follows from this lemma's assumptions. 
We continue to asses the size $\frac{|\Gamma_p| R_{\psi_\alpha}^2}{\sigma^2}$. We have that
\begin{eqnarray}
        \frac{|\Gamma_p| R_{\psi_\alpha}^2}{\sigma^2} &\lesssim& \frac{Q 4^{-2p} \frac{NK\mu^2 \mu_H^2}{Q^2}}{ \frac{16^{-p}}{Q} \max (SK\mu^2, N\mu_H^2)} \\ \nonumber
        &\lesssim&  \frac{K\mu^2 N\mu_H^2}{ N\mu_H^2}
        = K\mu^2\leq L.
        % \lesssim \max (K\mu^2, N\mu_H^2)
\end{eqnarray}
Finally, we can set these sizes in Lemma \ref{Matrix_Bernstein_Inequality} with $t = (\omega + 1)\log L$, $\alpha = 1$ and get with probability $1-\mathcal{O}(L^{-\omega-1})$,
% \begin{equation}
\begin{eqnarray}
    \norm{\sum_{l\in\Gamma_p} Z_l} \lesssim_\omega 4^{-p}
      \max \bigg{\{} &\sqrt{\frac{  (SK\mu^2+ N\mu_H^2)\log L}{Q}},\\ \nonumber
     &\frac{K\mu^2+ N\mu_H^2}{Q}\log^2 L \bigg{\}},
\end{eqnarray}
% \end{equation}
for a $p\in[P]$. Taking the union bound for all $p\in[P]$, we get $\norm{ \frac{L}{Q} (\mathcal{A}^p)^* \mathcal{A}^p S^p W_{p-1} - W_{p-1}} \leq \frac{1}{4^{p+1}}$ with prob. of at least $1-P\mathcal{O}(L^{-\omega-1}) = 1- \mathcal{O}(L^{-\omega})$, which finishes the proof.
\end{proof}
It thus remains to prove that $\mu_p \leq \frac{1}{4} \mu_{p-1}$.
\begin{lemma}\label{lemma:mu_p}
Let $\omega \geq 1$. if 
\begin{equation}
     Q \gtrsim S \max(K\mu^2 , N\mu_H^2)\log^2 L
\end{equation}
then with probability $1-\mathcal{O}(L^{-\omega})$ it holds that $\mu_p \leq \frac{1}{4} \mu_{p-1}$ for all $p\in [P-1]$.
\end{lemma}
\begin{proof}
Recall that by the definition of $\mu_p$ in \eqref{mu_p}, we need to show that for all $l\in \Gamma_p$ and for all $p\in[P-1]$,
\begin{equation}\label{mu_p_goal}
    \sqrt{L} \norm{W^*_p S_{p+1} b_l}_2 \leq \frac{1}{4} \mu_{p-1}.
\end{equation}
By \eqref{W_p_rec}, we have that $ W_p = W_{p-1} - \frac{L}{Q} \mathcal{P}_\mathcal{T} (\mathcal{A}^p)^* \mathcal{A}^p S^p W_{p-1}.$
% \begin{equation}
%     W_p = W_{p-1} - \frac{L}{Q} \mathcal{P}_\mathcal{T} (\mathcal{A}^p)^* \mathcal{A}^p S^p W_{p-1}.
% \end{equation}
Thus, we can use \eqref{P_T}, \eqref{vec_proj} together with \eqref{W_p_before_proj} to write
\begin{eqnarray}
    W_p = \frac{L}{Q}  \sum_{j\in\Gamma_p}(&\hat{H}\hat{H}^* b_j w_j^* (I - c_j c_j^*) +\\ \nonumber
                                            &(I - \hat{H}\hat{H}^*) b_j w_j^* (I - c_j c_j^*) \hat{M}\hat{M}^*\ ),
\end{eqnarray}
where $w_j = W^*_{p-1} S_p b_j$ as in \eqref{w_l} and  $W_{p-1} \in \mathcal{T}$. Using the triangle inequality, we have that
\begin{eqnarray}\label{two_summands}
     \nonumber && \hspace{-0.3in}\norm{W^*_p S_{p+1} b_l}_2 \leq \norm{\frac{L}{Q} \sum_{j\in\Gamma_p} (I - c_j c_j^*) w_j b_j^* \hat{H}\hat{H}^*
                                            S_{p+1} b_l}_2  + \\ \nonumber
                        &&\norm{\frac{L}{Q} \sum_{j\in\Gamma_p} \hat{M}\hat{M}^* (I - c_j c_j^*) w_j b_j^* (I - \hat{H}\hat{H}^*) S_{p+1} b_l}_2 \\
                            &&  \triangleq \norm{\sum_{j\in\Gamma_p} u_j}_2 + \norm{\sum_{j\in\Gamma_p} v_j}_2
\end{eqnarray}
We now use Lemma \ref{Matrix_Bernstein_Inequality} again to bound the two summands. For the first part, we have 
\begin{eqnarray}
    \norm{\sum_{j\in\Gamma_p} E[u_j u_j^*] } 
    &=& \frac{L^2}{Q^2}  \norm{\sum_{j\in\Gamma_p} w_j b_j^* \hat{H}\hat{H}^* S_{p+1} b_l }^2_2 \\ \nonumber
    &\leq& \frac{L^2}{Q^2}  \sum_{j\in\Gamma_p, s\in[S]} \norm{w_j b_j^* \hat{h}_s\hat{h}_s^* S_{p+1} b_l }^2_2 \\ \nonumber
    &=& \frac{L^2}{Q^2} \sum_{s\in[S]} |\hat{h}_s^* S_{p+1} b_l|^2 \sum_{j\in\Gamma_p} |b_j^* \hat{h}_s|^2 \norm{w_j}_2^2 \\ \nonumber
    &\leq& \frac{\mu_H^2 \mu_{p-1}^2}{QL} \sum_{s\in[S]} \norm{T_p^{1/2}\hat{h}_s}_2^2 
    \lesssim \frac{S \mu_H^2 \mu_{p-1}^2}{QL},
\end{eqnarray}
where we have used Lemma 12 in\cite{subspace_prior_blind_deconv} again and by \eqref{eq:Tp^1/2} and the definitions in \eqref{H_incoherence}, \eqref{mu_p}, \eqref{T_p} and \eqref{vec_proj}. Similarly, 
\begin{eqnarray}
   \nonumber \norm{\sum_{j\in\Gamma_p} E[u_j^* u_j] } 
    &\leq& \frac{L^2N}{Q^2}  \sum_{j\in\Gamma_p, s\in[S]} \norm{w_j b_j^* \hat{h}_s\hat{h}_s^* S_{p+1} b_l }^2_2 \\ 
    &\lesssim& \frac{S N \mu_H^2 \mu_{p-1}^2}{QL},
\end{eqnarray}
again using Lemma 11 in\cite{subspace_prior_blind_deconv}, leading to $\sigma^2 \lesssim \frac{S N \mu_H^2 \mu_{p-1}^2}{QL}$. Furthermore,
\begin{eqnarray}
    R_{\psi_1} &=& 
    \underset{j\in\Gamma_p}{\max} \norm{\norm{ \frac{L}{Q}(I - c_j c_j^*) w_j b_j^* \hat{H}\hat{H}^* S_{p+1} b_l}_2}_{\psi_1} \\ \nonumber
    &=& \frac{L}{Q} \underset{j\in\Gamma_p}{\max} \norm{\norm{ \sum_{s\in[S]} (I - c_j c_j^*) w_j b_j^* \hat{h}_s\hat{h}_s^* S_{p+1} b_l}_2}_{\psi_1} \\\nonumber
    &\leq& \frac{L}{Q} \frac{\mu_H^2}{L} S \ \underset{j\in\Gamma_p}{\max} \norm{\norm{ (I - c_j c_j^*) w_j}_2}_{\psi_1}\\\nonumber
    &\lesssim& \frac{L}{Q} \frac{\mu_H^2}{L} S \ \underset{j\in\Gamma_p}{\max}\sqrt{N}\norm{w_j}_2
    \leq\frac{S\sqrt{N}\mu_H^2 \mu_{p-1}}{Q\sqrt{L}}.
    % \\
    % &\leq& \frac{L}{Q} \frac{\mu_H^2}{L} S \sqrt{N} \frac{\mu_{p-1}}{\sqrt{L}} = 
    % \frac{S\sqrt{N}\mu_H^2 \mu_{p-1}}{Q\sqrt{L}}.
\end{eqnarray}
Now, we can assess
    \begin{eqnarray}
        \frac{|\Gamma_p| R_{\psi_1}^2}{\sigma^2} &\lesssim Q \frac{S^2 N \mu_H^4 \mu_{p-1}^2}{Q^2 L}
        \frac{QL}{S N \mu_H^2 \mu_{p-1}^2} 
         =S \mu_H^2
         \lesssim S L.
    \end{eqnarray}
Finally, we can write with $t = (\omega + 1)\log L$, $\alpha = 1$
\begin{eqnarray}
        &\norm{\frac{L}{Q} \sum_{j\in\Gamma_p} (I - c_j c_j^*) w_j b_j^* \hat{H}\hat{H}^*S_{p+1} b_l}_2  \lesssim \\ \nonumber
        &\frac{\mu_{p-1}}{\sqrt{L}} \max \bigg{\{} \sqrt{\frac{SN\mu_H^2}{Q} \log L} ,
        \frac{S\sqrt{N}\mu_H^2}{Q} \log^2 L \bigg{\}}
\end{eqnarray}
with a probability of at least $1-\mathcal{O}(L^{-\omega-1})$ for a fixed $p\in[P]$. Taking the union bound for all $p\in[P]$, we get that if $Q \gtrsim SN\mu_H^2 \log^2 L$, then the first summand in \eqref{two_summands} is bounded as required by \eqref{mu_p_goal} with probability of at least $1-P\mathcal{O}(L^{-\omega-1}) = 1- \mathcal{O}(L^{-\omega})$.

We now repeat the same process for the second summand in \eqref{two_summands}. We use the $Z_j$ notation defined in \eqref{Z_l}.
\begin{eqnarray}
    &&\nonumber \norm{\sum_{j\in\Gamma_p} E[v_j v_j^*] }  \leq 
    \sum_{j\in\Gamma_p} E \lVert  \hat{M}\hat{M}^* Z_j^* (I-\hat{H}\hat{H}^*) S_{p+1} b_l\ \cdot\\\nonumber
    && \qquad \qquad \qquad \qquad \qquad \ \ b_l^* S_{p+1} (I-\hat{H}\hat{H}^*) Z_j \hat{M}\hat{M}^* \rVert \\\nonumber
    && \leq \sum_{j\in\Gamma_p} E \lVert Z_j^* (I-\hat{H}\hat{H}^*) S_{p+1} b_l b_l^* S_{p+1} (I-\hat{H}\hat{H}^*) Z_j \rVert \\\nonumber
    && =\frac{L^2}{Q^2} \sum_{j\in\Gamma_p} E \lVert (I - c_j c_j^*) w_j b_j^* (I-\hat{H}\hat{H}^*) S_{p+1} b_l\ \cdot\\ \nonumber
    && \qquad \qquad \ \ \  b_l^* S_{p+1} (I-\hat{H}\hat{H}^*) b_j w_j^* (I - c_j c_j^*)\rVert \\\nonumber
    &&=\frac{L^2}{Q^2}  \sum_{j\in\Gamma_p} \norm{w_j b_j^* (I-\hat{H}\hat{H}^*) S_{p+1} b_l }^2_2 \\
    && \leq \frac{L^2}{Q^2}  \sum_{j\in\Gamma_p} \norm{w_j}^2_2 |b_j^* (I-\hat{H}\hat{H}^*) S_{p+1} b_l |^2 \\\nonumber
    &&\leq \frac{L^2}{Q^2} \frac{\mu_{p-1}^2}{L} \frac{Q}{L} \norm{T_p^{1/2} (I-\hat{H}\hat{H}^*) S_{p+1} b_l}^2  \\ \nonumber
    &&\leq\frac{L^2}{Q^2} \frac{\mu_{p-1}^2}{L} \frac{Q}{L} \norm{T_p^{1/2}}^2 \norm{(I-\hat{H}\hat{H}^*)}^2 \norm{S_{p+1}}^2 \norm{b_l}^2 \\ \nonumber
    && \lesssim \frac{L^2}{Q^2} \frac{\mu_{p-1}^2}{L} \frac{Q}{L} \norm{b_l}^2 \leq 
      \frac{L^2}{Q^2} \frac{\mu_{p-1}^2}{L} \frac{Q}{L} \frac{K\mu^2}{L} =
     \frac{K\mu^2 \mu_{p-1}^2}{QL}
\end{eqnarray}
where we have used Lemma 12 in\cite{subspace_prior_blind_deconv} again and by \eqref{eq:Tp^1/2} and the definitions in \eqref{H_incoherence}, \eqref{mu_p}, \eqref{T_p} and \eqref{vec_proj}. 

Similarly, 
\begin{eqnarray}
    &&\norm{\sum_{j\in\Gamma_p} E[v_j^* v_j] }   \\ \nonumber
    && \leq \frac{L^2}{Q^2} \sum_{j\in\Gamma_p} E \lVert b_l^* S_{p+1} (I-\hat{H}\hat{H}^*) b_j w_j^* (I - c_j c_j^*)
    \hat{M}\hat{M}^* \cdot\\ \nonumber
    && \qquad \qquad \qquad \ \ \  (I - c_j c_j^*) w_j b_j^* (I-\hat{H}\hat{H}^*) S_{p+1} b_l\rVert \\ \nonumber
     && = \frac{L^2S}{Q^2}  \sum_{j\in\Gamma_p} \norm{w_j b_j^* (I-\hat{H}\hat{H}^*) S_{p+1} b_l }^2_2 \lesssim 
     \frac{SK\mu^2 \mu_{p-1}^2}{QL}
\end{eqnarray}
again using Lemma 12 in\cite{subspace_prior_blind_deconv}. Thus, $\sigma^2 \lesssim \frac{SK\mu^2 \mu_{p-1}^2}{QL}$. Furthermore,
\begin{flalign}
  \nonumber  &R_{\psi_1} =
    \underset{j\in\Gamma_p}{\max} \norm{\norm{ \frac{L}{Q}\hat{M}\hat{M}^*(I - c_j c_j^*) w_j b_j^* (I-\hat{H}\hat{H}^*) S_{p+1} b_l}_2}_{\psi_1} \\ \nonumber
    &=\underset{j\in\Gamma_p}{\max} \norm{\norm{ \frac{L}{Q}\hat{M}^*(I - c_j c_j^*) w_j b_j^* (I-\hat{H}\hat{H}^*) S_{p+1} b_l}_2}_{\psi_1} \\ \nonumber
    &\leq \frac{L}{Q} \underset{j\in\Gamma_p}{\max} |b_j^*(I-\hat{H}\hat{H}^*) S_{p+1} b_l| \sum_{s\in[S]}\norm{ \hat{m}_s^*(I - c_j c_j^*) w_j}_{\psi_1} \\ 
    &\leq \frac{L}{Q}  \frac{K\mu^2}{L} \sum_{s\in[S]}\norm{\hat{m}_s^*}_2 \underset{j\in\Gamma_p}{\max} \norm{w_j}_2
    \leq  \frac{SK\mu^2\mu_{p-1}}{Q\sqrt{L}},
\end{flalign}
where we have used Lemma 39 in \cite{linear_blind_deconv_demix_one_measurement}. Now, we can assess
\begin{eqnarray}
        \frac{|\Gamma_p| R_{\psi_1}^2}{\sigma^2} \lesssim Q \frac{S^2K^2\mu^4\mu^2_{p-1}}{Q^2 L}
        \frac{QL}{SK \mu^2 \mu_{p-1}^2} 
        =SK\mu^2 \leq SL.
\end{eqnarray}
Finally, we can write with $t = (\omega + 1)\log L$, $\alpha = 1$
\begin{eqnarray}
     \nonumber   &\norm{\frac{L}{Q} \sum_{j\in\Gamma_p} \hat{M}\hat{M}^* (I - c_j c_j^*) w_j b_j^* (I - \hat{H}\hat{H}^*) S_{p+1} b_l}_2   \\
        & \lesssim \frac{\mu_{p-1}}{\sqrt{L}} \max \bigg{\{}\sqrt{\frac{S K\mu^2 \log L}{Q}},
        \frac{SK\mu^2}{Q} \log^2 L \bigg{\}},
\end{eqnarray}
with a probability of at least $1-\mathcal{O}(L^{-\omega-1})$ for a fixed $p\in[P]$. Taking the union bound for all $p\in[P]$, we get that if $Q \gtrsim SK \mu^2 \log^2 L$, then the second summand in \eqref{two_summands} is bounded as required by \eqref{mu_p_goal} with probability of at least $1-P\mathcal{O}(L^{-\omega-1}) = 1- \mathcal{O}(L^{-\omega})$, which finishes the proof.
\end{proof}

\subsection{Proof of Lemma \ref{lemma:suprema_of_chaos_bounds}}\label{proof:suprema_of_chaos_bounds}
\begin{proof}[Proof of Lemma \ref{lemma:suprema_of_chaos_bounds}]
To prove this lemma, we use the following two technical lemmas. The proof of the first appears in Appendix~\ref{proof:covering_B_M} and of the second in \cite{linear_blind_deconv_demix_one_measurement}. First, let us denote the unit ball with respect to $\norm{\cdot}_2$ by $B(0,1)$ (throughout the paper).

\begin{lemma}\label{lemma:covering_B_M}
Let $\mathcal{B}^M$ be defined by \eqref{eq:ball_definitions}. Then
\begin{eqnarray}\label{RHS}
        N(\mathcal{B}^M, \norm{\cdot}_B, \epsilon) \leq 
        N \Big{(} B(0,1) \subset \mathbb{R}^S, \norm{\cdot}_2, \frac{\epsilon}{2\sqrt{K\mu}} \Big{)} \\ \nonumber
        \cdot N^S \Big{(} B(0,1) \subset \mathbb{C}^{K}, \norm{\cdot}_B, \frac{\epsilon}{2} \Big{)}
\end{eqnarray}
\end{lemma}
\begin{lemma}\label{lem:log_covering_number}
(A private case of Lemma 27 in \cite{linear_blind_deconv_demix_one_measurement}). 
\begin{equation}
    \log N \Big{(} B(0,1) \subset \mathbb{C}^K, \norm{\cdot}_B, \frac{\epsilon}{2} \Big{)} \lesssim
    \frac{K \mu^2}{\epsilon^2} \log L.
\end{equation}
\end{lemma}
We now turn to prove Lemma~\ref{lemma:suprema_of_chaos_bounds}. The first inequality \eqref{eq:ineq_1} follows the fact that
\begin{equation}
    % \begin{split}
        d_F(\mathcal{X}) = 
        \underset{X \in \mathcal{X}}{\sup} \norm{X}_F \leq 
        \underset{X \in \mathcal{W}^p}{\sup} \norm{X}_F  \leq 3,
    % \end{split}
\end{equation}
where the last inequality holds since $\mathcal{W}^p = \mathcal{B}^M + \mathcal{B}^H + \mathcal{B}^{S^p H}$ and all the elements in these sets are normalized. 

To prove the second inequality \eqref{eq:ineq_2}, we use the definitions of $d_B(\mathcal{X})$ and $\norm{\cdot}_B$, to get 
\begin{eqnarray}
       & d_B(\mathcal{X}) = 
        \underset{X \in \mathcal{X}}{\sup} \norm{X}_B = 
        \underset{X \in \mathcal{X}}{\sup} \sqrt{L} \ \underset{l \in [L]}{\max} \norm{X^* b_l}_F \leq \\ 
        &\underset{X \in \mathcal{X}}{\sup} \sqrt{L} \norm{X}_F \underset{l \in [L]}{\max} \norm{b_l}_2 \leq 
        \underset{X \in \mathcal{X}}{\sup} \norm{X}_F \sqrt{K} \mu \leq
        3 \sqrt{K} \mu, \nonumber
\end{eqnarray}
where the first inequality is due to the Frobenius norm properties, the second is due to the definition of $\mu$ in \eqref{B_incoherence} and the last is due to \eqref{eq:ineq_1}.

For \eqref{eq:ineq_3}, we can use Lemma 12 in \cite{linear_blind_deconv_demix_one_measurement} to obtain
\begin{flalign}
        &\gamma_2(\mathcal{W}^p, \norm{\cdot}_B)  \lesssim \\
        &\gamma_2(\mathcal{B}^H, \norm{\cdot}_B) + 
        \gamma_2(\mathcal{B}^M, \norm{\cdot}_B) +
        \gamma_2(\mathcal{B}^{S^p H}, \norm{\cdot}_B), \nonumber
\end{flalign}
where $\gamma_2(\mathcal{W}, \norm{\cdot}_B)$ is bounded analogously. 

First we bound $\gamma_2(\mathcal{B}^H, \norm{\cdot}_B)$. Let $\textbf{U}=\hat{H}U^*, \textbf{V}=\hat{H}V^* \in \mathcal{B}^M$. Then 
\begin{flalign}
\label{eq:UV_B_mu_H_inequality}
       & \norm{\textbf{U} - \textbf{V}}^2_B =
         \norm{\hat{H}(U^* - V^*)}^2_B  = \\ \nonumber
       &=  L \underset{l \in [L]}{\max}\norm{(U- V)\hat{H}^* b_l}^2_2 
        \leq L \underset{l \in [L]}{\max} \sum_{s \in [S]} \norm{\hat{h}_s^*b_l(u_s - v_s)}^2_2 \\ \nonumber
       &= L \underset{l \in [L]}{\max} \sum_{s \in [S]} | \hat{h}_s^*b_l |^2 \norm{(u_s - v_s)}^2_2  \leq 
        \mu_H^2 \sum_{s \in [S]} \norm{(u_s - v_s)}^2_2 \\ \nonumber
       &= \mu_H^2 \norm{(U - V)}^2_F  = 
         \mu_H^2 \norm{\textbf{U} - \textbf{V}}^2_F 
\end{flalign}
where the inequality is due to \eqref{H_incoherence} and the last equality holds because the Frobenius norm is unitary invariant. 
Using \eqref{eq:UV_B_mu_H_inequality} followed by Dudley's inequality (Th. \ref{theorem:Dudley's_ineq}), implies 
\begin{flalign}
\label{eq:gamma2_BH_Dudley}
        &\gamma_2(\mathcal{B}^H, \norm{\cdot}_B)  \leq \mu_H \gamma_2(\mathcal{B}^H, \norm{\cdot}_F) \lesssim \\
        & \mu_H \int_0^1 \sqrt{\log\ N(\mathcal{B}^H, \norm{\cdot}_F, \epsilon)} d\epsilon 
         \lesssim \mu_H \sqrt{SN}, \nonumber
\end{flalign}
where the last inequality holds since $(\mathcal{B}^H, \norm{\cdot}_F)$ is isometric to $(B(0,1) \subset\mathbb{R}^{2SN}, \norm{\cdot}_2)$ and a standard volumetric estimate.

Similarly, let $\textbf{U}=S_p \hat{H}U^*, \textbf{V}=S_p \hat{H}V^* \in \mathcal{B}^{S^pH}$. Then 
\begin{flalign}
        & \norm{\textbf{U} - \textbf{V}}^2_B = 
        \norm{S_p \hat{H}(U^* - V^*)}^2_B  = \\ \nonumber
        & L \underset{l \in [L]}{\max}\norm{(U- V)\hat{H}^* S_p b_l}^2_2  = 
         L \underset{l \in [L]}{\max} \norm{  \sum_{s \in [S]}\hat{h}_s^* S_p b_l(u_s - v_s)}^2_2  \\ \nonumber
        & \leq L \underset{l \in [L]}{\max} \sum_{s \in [S]} | \hat{h}_s^* S_p b_l |^2 \norm{u_s - v_s}^2_2  \leq 
         \mu_H^2 \sum_{s \in [S]} \norm{u_s - v_s}^2_2   \\ \nonumber
        &= \mu_H^2 \norm{U-V}_F^2 = 
         \mu_H^2 \norm{\hat{H}(U^*-V^*)}_F^2  \\ \nonumber
         &=\mu_H^2 \norm{T_p S_p\hat{H}(U^*-V^*)}_F^2,
\end{flalign}  
where the last equality holds since $T_p S_p = I$ and the rest is as in \eqref{eq:UV_B_mu_H_inequality}.
The final norm can be split into 
\begin{flalign}
   \nonumber     & \mu_H^2 \norm{T_p S_p\hat{H}(U^*-V^*)}_F^2 \leq 
         \mu_H^2 \norm{T_p}^2 \norm{S_p\hat{H}(U^*-V^*)}_F^2  \\
        &\leq \mu_H^2  (1 + \nu)^2  \norm{\textbf{U} - \textbf{V}}_F^2 \lesssim 
         \mu_H^2   \norm{\textbf{U} - \textbf{V}}_F^2,
\end{flalign}
where the second inequality  is due to $\norm{T_p} \leq 1 + \nu$. 
Using Dudley's inequality as in \eqref{eq:gamma2_BH_Dudley}, we get 
\begin{equation}\label{eq:bound_B_SpH}
    \gamma_2(\mathcal{B}^{S^pH}, \norm{\cdot}_B) \lesssim \mu_H \sqrt{SN}. 
\end{equation}

To bound $\gamma_2(\mathcal{B}^M, \norm{\cdot}_B)$, notice that $d_B(\mathcal{B}^M) \leq \sqrt{K}\mu$, so by Dudley's inequality we get
\begin{equation}
    \gamma_2(\mathcal{B}^M, \norm{\cdot}_B) \lesssim \int_0^{\sqrt{K}\mu} \sqrt{\log\ N(\mathcal{B}^M, \norm{\cdot}_B, \epsilon)} d\epsilon.
\end{equation} 
For the rhs, we use Lemma \ref{lemma:covering_B_M} and have
\begin{equation}\label{eq:integrals_to_cover_Bm}
    \begin{split}
        & \gamma_2(\mathcal{B}^M, \norm{\cdot}_B) \lesssim \\
        & \int_0^{\sqrt{K}\mu} \sqrt{\log\ N \bigg{(} B(0,1) \subset \mathbb{R}^S, \norm{\cdot}_2,                              \frac{\epsilon}{2\sqrt{K}\mu} \bigg{)} } d\epsilon  + \\
        & \int_0^{\sqrt{K}\mu} \sqrt{ S \ \log\ N \bigg{(} B(0,1) \subset \mathbb{C}^K, \norm{\cdot}_B,                              \frac{\epsilon}{2} \bigg{)} } d\epsilon
    \end{split}
\end{equation}
% These integrals can be bound in a similar way to step 3 in the proof of Lemma 28 in \cite{linear_blind_deconv_demix_one_measurement}. 
Thus, the first integral is bounded by
\begin{equation}\label{eq:first_int_result}
\begin{split}
    & \int_0^{\sqrt{K}\mu} \sqrt{\log\ N \bigg{(} B(0,1) \subset \mathbb{R}^S, \norm{\cdot}_2,                              \frac{\epsilon}{2\sqrt{K}\mu} \bigg{)} } d\epsilon \leq \\
    & \sqrt{S} \int_0^{\sqrt{K}\mu} \sqrt{\log\bigg{(} 1 + \frac{4\sqrt{K} \mu}{\epsilon} \bigg{)} } d\epsilon \lesssim \sqrt{SK}\mu,
\end{split}
\end{equation}
where we have used a standard volumetric estimate and a change of variables. For the second integral in \eqref{eq:integrals_to_cover_Bm}, we provide a private case of the derivation in \cite{linear_blind_deconv_demix_one_measurement} for completeness. We now split the second integral in \eqref{eq:integrals_to_cover_Bm} to two integration intervals: $[0,1]$ and $[1, \sqrt{K}\mu]$. For $\epsilon \in (0,1)$, we define
\begin{eqnarray}
        B(0,1) \subset  \sqrt{K}\mu B_{\norm{\cdot}_B}(0,1) \triangleq \{ x \in \mathbb{C}^K | \norm{x}_B \leq \sqrt{K}\mu \}.
\end{eqnarray}
This implies that
\begin{flalign}
\label{eq:B_ball_B_norm_covering_bound}
        &N (B(0,1) \subset  \mathbb{C}^K, \norm{\cdot}_B,\epsilon) \leq \\ \nonumber
        & N \bigg{(} B(0,1)_{\norm{\cdot}_B} \subset \mathbb{C}^K, \norm{\cdot}_B,\frac{\epsilon}{\sqrt{K}\mu} \bigg{)} \leq
         \bigg{(} 1 + \frac{2\sqrt{K}\mu}{\epsilon} \bigg{)}^{2K},
\end{flalign}
where the last inequality is a standard bound for the covering number.
For the interval $[0,1]$ we get the following bound
\begin{flalign}\label{eq:first_interval}
        & \int_0^1 \sqrt{S \log N \bigg{(} B(0,1)\subset  \mathbb{C}^K, \norm{\cdot}_B,\frac{\epsilon}{2} \bigg{)}} d\epsilon  \\ \nonumber
        &\leq \sqrt{2KS} \int_0^1 \sqrt{ \log \bigg{(} 1 + \frac{2\sqrt{K}\mu}{\epsilon} \bigg{)}} d\epsilon  \\
        &\leq \sqrt{2KS\log ( e( 1 + 2\sqrt{K}\mu ) ) }, \nonumber
\end{flalign}
where the first inequality is due to \eqref{eq:B_ball_B_norm_covering_bound} and the second one is due to Lemma C.9 in \cite{math_intro_CS}.
Now we deal with the case where $\epsilon \in (1, \sqrt{K}\mu)$. Using Lemma \ref{lem:log_covering_number}, we get
\begin{flalign}\label{eq:second_interval}
        & \int_1^{\sqrt{K}\mu} \sqrt{S N \bigg{(} B(0,1) \subset  \mathbb{C}^K, \norm{\cdot}_B,\frac{\epsilon}{2} \bigg{)}} d\epsilon \lesssim \\ \nonumber
        & \int_1^{\sqrt{K}\mu} \frac{\sqrt{SK\log(L)}\mu}{\epsilon} d\epsilon \lesssim  \sqrt{SK\log(L)}\mu \log(K\mu^2).
\end{flalign}
Combining \eqref{eq:first_interval} with \eqref{eq:second_interval} provides us with 
\begin{equation}\label{eq:second_int_result}
    \begin{split}
        & \int_0^{\sqrt{K}\mu} \sqrt{ S \ \log\ N \bigg{(} B(0,1) \subset \mathbb{C}^K, \norm{\cdot}_B, \frac{\epsilon}{2} \bigg{)} } d\epsilon \lesssim \\
        & \sqrt{SK\log(L)}\mu \log(K\mu^2),
    \end{split}
\end{equation}
where we use the fact that \eqref{eq:second_interval} is the dominant interval. Plugging \eqref{eq:first_int_result} and \eqref{eq:second_int_result} in \eqref{eq:integrals_to_cover_Bm}, and considering again the dominant part, leads to
\begin{equation}\label{eq:bound_B_M}
    \gamma_2(\mathcal{B}^M, \norm{\cdot}_B) \lesssim  \sqrt{SK\log(L)}\mu \log(K\mu^2).
\end{equation}
The result stated in \eqref{eq:ineq_3} is given by the summation of the three bounds in \eqref{eq:gamma2_BH_Dudley}, \eqref{eq:bound_B_SpH} and \eqref{eq:bound_B_M}.
\end{proof}

\subsection{Proof of Lemma \ref{lemma:covering_B_M}}\label{proof:covering_B_M}
\begin{proof}[Proof of Lemma \ref{lemma:covering_B_M}]
For all $s\in[S]$, let $\mathcal{N}_s$ be an $\frac{\epsilon}{2}$-cover of $B(0,1) \subset \mathbb{C}^{K}$ with respect to the $\norm{\cdot}_B$-norm and $\mathcal{O}$ be an $\frac{\epsilon}{2\sqrt{K\mu}}$-cover of $B(0,1) \subset \mathbb{R}^{S}$ with respect to the $\norm{\cdot}_2$ norm. We will show now that any $X = U\hat{M}^* \in \mathcal{B}^M$ can be approximated by $Y = \underset{s\in[S]}{\sum}\sigma_s v_s \hat{m}_s^*$ where $\sigma \in \mathcal{O}$ and $v_s \in \mathcal{N}_s$. Notice that the number of such $Y$s is bounded by the right hand side of the inequality in \eqref{RHS}. Thus, it remains to show that such a construction is possible.
% First, notice that
% \begin{eqnarray}
%         & \norm{A}_B^2 = 
%         \underset{l\in [L]}{\max} \norm{A^*b_l}_2^2 =  \underset{l\in [L]}{\max} \sum_{s\in[S]} \norm{a_s^*b_l}_2^2 \\ \nonumber
%         & \le \sum_{s\in[S]} \underset{l\in [L]}{\max} \norm{a_s^*b_l}_2^2 =
%         \sum_{s\in[S]} \norm{a_s}_B^2
% \end{eqnarray}
% where $A\in \mathbb{C}^{K\times S}$. This infers that the matrix B-norm ($\norm{A}_B$) in a space of matrices is equivalent to the vector B-norm ($\norm{a}_B$) in a space of vectors. Thus, %considering the volume $B(0,1) \subset \mathbb{C}^K$,
% \begin{eqnarray}
%     N(B(0,1) &\subset& \mathbb{C}^{K\times S}, \norm{A}_B, \epsilon) =\\ N^S(B(0,1) &\subset& \mathbb{C}^{K}, \norm{a}_B, \epsilon). \nonumber
% \end{eqnarray}
Since $\sigma \in \mathcal{O}$, we may pick it to satisfy
\begin{equation}
    \sqrt{\sum_{s \in [S]} (\norm{u_s}_2 - \sigma_{s})^2} \leq \frac{\epsilon}{2\sqrt{K}\mu}.
\end{equation}
Notice, that $(\norm{u_1}_2, \dots, \norm{u_S}_2) \in B(0,1)$, since $X \in \mathcal{B}^M$ and $\hat{M}$ is orthonormal. In a similar way, since  $v_s \in \mathcal{N}_s$, we select it such that
\begin{equation}\label{eq:choose_V}
    \norm{\frac{1}{\norm{u_s}_2} u_s - v_s}_B \leq \frac{\epsilon}{2}
\end{equation}
for all $s\in S$. Thus, for $\hat{Y} = \underset{s\in[S]}{\sum} \norm{u_s}_2 v_s \hat{m}^*$, we have  
% \begin{flalign}
% \label{eq:X_Y_hat_inequality_4covering}
%         &\norm{X-\hat{Y}}^2_B \leq 
%         \sum_{s \in [S]} \norm{u_s \hat{m}_s^* - \norm{u_s}_2 v_s \hat{m}_s^*}^2_B  \\ \nonumber
%         &= \sum_{s \in [S]} \underset{l\in[L]}{\max}\norm{\hat{m}_s(u_s - \norm{u_s}_2 v_s)^* b_l}^2_2  \\ \nonumber
%         &= \sum_{s \in [S]} \underset{l\in[L]}{\max}\norm{\hat{m}_s}_2 |(u_s - \norm{u_s}_2 v_s)^* b_l|^2  \\ \nonumber
%         &=\sum_{s \in [S]} \norm{u_s - \norm{u_s}_2 v_s}^2_B = 
%         \sum_{s \in [S]} \norm{\norm{u_s}_2 \bigg{(} \frac{1}{\norm{u_s}_2}u_s - v_s\bigg{)}}^2_B \\ \nonumber
%         &\leq \frac{\epsilon^2}{4} \sum_{s \in [S]} \norm{u_s}_2^2 = 
%         \frac{\epsilon^2}{4} \norm{U}_F^2 = 
%         \frac{\epsilon^2}{4} \norm{X}_F^2 \leq
%         \frac{\epsilon^2}{4}.
% \end{flalign}
\begin{flalign}
\label{eq:X_Y_hat_inequality_4covering_1}
        &\norm{X-\hat{Y}}^2_B = 
        \norm{\sum_{s \in [S]} (u_s -\norm{u_s}_2 v_s) \hat{m}_s^*}^2_B  \\ \nonumber
        &= L \underset{l\in[L]}{\max} \norm{ \sum_{s \in [S]} \hat{m}_s(u_s - \norm{u_s}_2 v_s)^* b_l}^2_2  \\ \nonumber
        &= L \underset{l\in[L]}{\max}  \sum_{s,k \in [S]} b_l^* (u_s - \norm{u_s}_2 v_s) \hat{m}_s^* \hat{m}_k  (u_k - \norm{u_k}_2 v_k)^* b_l  \\ \nonumber
        &= L \underset{l\in[L]}{\max}  \sum_{s \in [S]} b_l^* (u_s - \norm{u_s}_2 v_s) (u_s - \norm{u_s}_2 v_s)^* b_l 
\end{flalign}
where the second equality is due to the definition of $\norm{\cdot}_B$ in \eqref{def:B_norm} combined with the non-negativity of the norm, and the last step follows the orthonormality of $\hat{M}$. Next, we continue to bound
\begin{flalign}
\label{eq:X_Y_hat_inequality_4covering_2}
        &L \underset{l\in[L]}{\max}  \sum_{s \in [S]} b_l^* (u_s - \norm{u_s}_2 v_s) (u_s - \norm{u_s}_2 v_s)^* b_l 
        \\ \nonumber
        &\leq \sum_{s \in [S]} L \underset{l\in[L]}{\max} (b_l^* (u_s - \norm{u_s}_2 v_s) (u_s - \norm{u_s}_2 v_s)^* b_l)\\ \nonumber
        &=\sum_{s \in [S]} \norm{u_s - \norm{u_s}_2 v_s}^2_B = 
        \sum_{s \in [S]} \norm{\norm{u_s}_2 \bigg{(} \frac{1}{\norm{u_s}_2}u_s - v_s\bigg{)}}^2_B \\ \nonumber
        &\leq \frac{\epsilon^2}{4} \sum_{s \in [S]} \norm{u_s}_2^2 = 
        \frac{\epsilon^2}{4} \norm{U}_F^2 
        \le
        \frac{\epsilon^2}{4} 
\end{flalign}
where the first equality is again due to due to the definition of $\norm{\cdot}_B$ in \eqref{def:B_norm} combined with the non-negativity of the norm, the second inequality is due to \eqref{eq:choose_V} and the last step holds since $X = U\hat{M}^*\in \mathcal{B}^M$ is normalized and $\hat{M}$ is orthonormal. To conclude this step, by \eqref{eq:X_Y_hat_inequality_4covering_1}, \eqref{eq:X_Y_hat_inequality_4covering_2} we have
\begin{equation}\label{eq:X_Y_hat_inequality_4covering}
    \norm{X-\hat{Y}}^2_B \leq \frac{\epsilon^2}{4}
\end{equation}
To complete the proof, we similarly have
% \begin{flalign}
% \label{eq:Y_Y_hat_inequality_4covering}
%         &\norm{\hat{Y} - Y}^2_B \leq  
%         \sum_{s \in [S]} \norm{(\norm{u_s}_2 - \sigma_{s}) v_s \hat{m}_s^*}^2_B = \\ \nonumber
%         &\sum_{s \in [S]} (\norm{u_s}_2 - \sigma_{s})^2 \norm{v_s}^2_B \leq 
%         \sum_{s \in [S]} K \mu^2 (\norm{u_s}_2 - \sigma_{s})^2  \leq  
%         \frac{\epsilon^2}{4}.
% \end{flalign}
\begin{flalign}
\label{eq:Y_Y_hat_inequality_4covering}
        &\norm{\hat{Y} - Y}^2_B =  
         \norm{\sum_{s \in [S]} (\norm{u_s}_2 - \sigma_{s}) v_s \hat{m}_s^*}^2_B \leq \\ \nonumber
        & \sum_{s \in [S]} \norm{(\norm{u_s}_2 - \sigma_{s}) v_s}_B^2 =
        \sum_{s \in [S]} (\norm{u_s}_2 - \sigma_{s})^2 \norm{v_s}^2_B \leq \\ \nonumber
        &\sum_{s \in [S]} K \mu^2 (\norm{u_s}_2 - \sigma_{s})^2  \leq  
        \frac{\epsilon^2}{4},
\end{flalign}
where again we used the orthonormlity of $\hat{M}$, the non-negativity of the norm and the fact that
\begin{flalign}
    \norm{v_s}_B &= \sqrt{L} \underset{l \in L}{\max} |v_s^*b_l| \\ \nonumber
    &\leq  \sqrt{L} \norm{v_s}_2 \underset{l \in L}{\max} \norm{b_l}_2 \leq \sqrt{K} \mu,    
\end{flalign}
which holds since $v_s \in \mathcal{N}_s$. Finally, by combining \eqref{eq:X_Y_hat_inequality_4covering} and \eqref{eq:Y_Y_hat_inequality_4covering}, we get $\norm{X - Y}_B \leq \epsilon$.
\end{proof}

\section*{Acknowledgment}
We thank the anonymous reviewers for their useful comments that helped to improve the paper. This research was supported by ERC-StG grant no. 757497 (SPADE).

\bibliographystyle{IEEEtran}
\bibliography{bibtex}
\end{document}